\documentclass[3p,times]{elsarticle}

\usepackage{ecrc}
\usepackage{aas_macros}

\volume{00}

\firstpage{1}

\journalname{Procedia Computer Science}

\runauth{}

\jid{procs}

\jnltitlelogo{Astronomy and Computing}

\usepackage{amssymb}

\usepackage{anysize}
\marginsize{1.5cm}{1.5cm}{2cm}{2cm}

\usepackage[T1]{fontenc}
\usepackage{graphicx}
\usepackage[cmex10]{amsmath}
\usepackage{array}
\usepackage{url}
\usepackage{amsfonts}
\usepackage{lmodern}
\usepackage{algpseudocode}
\usepackage{algorithmicx}
\usepackage{algorithm}
\usepackage{xfrac}
\usepackage{float}
\usepackage{multirow}
\usepackage{balance}
\usepackage{tikz}
\usepackage[pdftex,colorlinks=true,breaklinks=true,citecolor=blue,draft=false]{hyperref}
\usetikzlibrary{arrows,shapes,automata}
\tikzstyle{every picture}+=[remember picture]

\usepackage[figuresright]{rotating}

\DeclareMathOperator{\soft}{\op{S}}
\DeclareMathOperator{\identity}{\op{I}}

\DeclareMathOperator{\proj}{\op{P}}
\DeclareMathOperator*{\argmin}{argmin}

\newcommand{\ds}{\displaystyle}

\newcommand{\bm}{\boldsymbol}

\newcommand{\bb}{\mathbb}
\newcommand{\mc}{\mathcal}

\newcommand{\ac}[1]{\uppercase{#1}}

\newcommand{\op}[1]{\boldsymbol{\mc{#1}}}

\algblock[Block]{Block}{EndBlock}
\algblockdefx[Block]{Block}{EndBlock}%
[1]{{#1}}%
[1]{{#1}}

\newcommand{\Given}[1]{\State{\bf given} {#1}}
\newcommand{\Func}[2]{\Block{{\bf function} \textsc{#1}}$\big($#2$\big)$}
\newcommand{\EndFunc}[1]{\EndBlock{{\bf return} #1}}
\newcommand{\FuncCall}[2]{\textsc{#1}$\big($#2$\big)$}
\newcommand{\RepeatFor}[1]{\Repeat {\bf~for} {#1}}

\newcommand{\ParFor}[1]{\Block{{#1} \bf do~in~parallel}}
\newcommand{\EndParFor}{\EndBlock{\bf end}}

\newcommand{\ignore}[1]{\color{magenta}\color{black}}

\newcommand{\ec}{\color{black}}

\begin{document}

\begin{frontmatter}
																																																																				
																																																																				
																																																																				
	\dochead{}
																																																																				
	\title{Distributed and parallel sparse convex optimization for radio\\ interferometry with PURIFY}
																																																																				
																																																																				
	\author[label1]{Luke Pratley}
	\author[label1]{Jason D.~McEwen}
	\author[label2]{Mayeul d'Avezac}
	\author[label1]{Xiaohao Cai}
	\author[label2]{David Perez-Suarez}
	\author[label2]{Ilektra Christidi}
	\author[label2]{Roland Guichard}
																																																																				
	\address[label1]{Mullard Space Science Laboratory (MSSL), University College London (UCL), Holmbury St Mary, Surrey RH5 6NT, UK}
	\address[label2]{Research Software Development Group, Research IT Services, University College London (UCL), London WC1E 6BT, UK}
																																																																				
	\begin{abstract}
		{
		Next generation radio interferometric telescopes are entering an era of big data with extremely large data sets. While these telescopes can observe the sky in higher sensitivity and resolution than before, computational challenges in image reconstruction need to be overcome to realize the potential of forthcoming telescopes.  New methods in sparse image reconstruction and convex optimization techniques (cf.\ compressive sensing) have shown to produce higher fidelity reconstructions of simulations and real observations than traditional methods. This article presents distributed and parallel algorithms and implementations to perform sparse image reconstruction, with significant practical considerations that are important for implementing these algorithms for Big Data. We benchmark the algorithms presented, showing that they are considerably faster than their serial equivalents. We then pre-sample gridding kernels to scale the distributed algorithms to larger data sizes, showing application times for 1 Gb to 2.4 Tb data sets over 25 to 100 nodes for up to 50 billion visibilities, and find that the run-times for the distributed algorithms range from 100 milliseconds to 3 minutes per iteration. This work presents an important step in working towards computationally scalable and efficient algorithms and implementations that are needed to image observations of both extended and compact sources from next generation radio interferometers such as the SKA. The algorithms are implemented in the latest versions of the SOPT (\url{https://github.com/astro-informatics/sopt}) and PURIFY (\url{https://github.com/astro-informatics/purify}) software packages {(Versions 3.1.0)}, which have been released alongside of this article.}
	\end{abstract}
																																																																				
																																																																				
\end{frontmatter}


\section{Introduction}
\label{sec:intro}
Since the 1940's radio interferometric telescopes have enabled observations of the radio sky at higher resolution and sensitivity than ever possible using a single dish telescope \cite{mcc47}. By combining single dish radio telescopes into an array, the pairs of antenna directly sample Fourier coefficients of the radio sky \cite{R&H1960}. The further apart the antenna, the higher the resolution, and the more antenna, the higher the sensitivity.

Next generation radio interferometric telescopes such as the Square Kilometre Array (SKA) will revolutionize astronomy. The high sensitivity and resolution of images from these telescopes are expected to enable new scientific discoveries in both cosmology and astrophysics \cite{bra15}. Among the revolutionary science goals, these radio interferometric telescopes are designed to detect the star formation during the Epoch of Re-ionization (EoR) \cite{koo15}, and increase our understanding of the formation and evolution of the galaxy cluster environment and magnetic fields \cite{fer15,mjh15}.

However, radio interferometric telescopes have a limited number of antenna, meaning they cannot complete the Fourier domain of the radio sky, and therefore image reconstruction techniques need to be employed to estimate the true sky brightness distribution. This includes traditional methods such as CLEAN \cite{hog74,rau11,pra16} and maximum entropy \citep{abl74,cor85}, and state-of-the-art deconvolution methods such as sparse regularization algorithms \citep{wia09a,wia09b,mce11,car14,dab15,gar15,gir15,ono16,pra18,pra18a,dab18,cai17a,cai17b,cai17c,aki17} (cf.\ compressive sensing). Furthermore, measurements from all pairs of antenna will generate an enormous amount of data, with measurements varying in time and observational radio frequency; errors in the instrumental response also present a big data challenge \cite{bro15}. For the SKA to achieve revolutionary science goals, it is critical to overcome revolutionary challenges in big data image reconstruction. In \cite{car14,ono16}, several algorithms from convex optimization were presented as tools for this big data challenge. {Where it is suggested that these} algorithms can be distributed on a computing cluster for big data image reconstruction, while providing accurate representations of the radio sky, motivating the developments in this work.

To bridge the gap between research in new distributed convex optimization algorithms and application of these algorithms to real observations, the software packages SOPT and PURIFY have been developed in C++ for use by astronomers \cite{car14,pra18}. SOPT hosts a collection of convex optimization algorithms that are used to solve inverse problems in signal processing. PURIFY hosts algorithms in radio interferometric imaging, and interfaces with SOPT to perform sparse image reconstruction\footnote{Such algorithms have been used in the wider framework of compressive sensing.} of radio interferometric observations. 

PURIFY has been applied to simulated and real interferometric observations, and has been shown to produce sensitive, accurate and high-resolution models of the radio sky \cite{car14,pra18,pra18a}. Additionally, unlike CLEAN reconstruction algorithms (a standard in radio astronomy) no restoration process is needed (e.g.\ combining of the reconstruction and residuals of the reconstruction as routinely performed with CLEAN), increasing the scientific merit of the reconstruction \cite{pra18}. PURIFY was recently used in \cite{pra18a} to test and develop a new method of {distributed} wide-field interferometric image reconstruction, where wide-field and non-coplanar effects were modeled for over 17.5 million measurements from a real observation from the SKA precursor the Murchison Wide-Field Array (MWA) \cite{tin13} using distributed convex optimization algorithms presented in this work. This method was recently extended to larger data sets from the MWA of over 100 million visibilities.
{
We present the first implementation of a distributed convex optimization algorithm that uses MPI to reconstruct images from radio interferometric telescopes, which is now implemented in PURIFY. We build on the general distributed framework proposed by \cite{ono16}, but develop a number of alternative algorithms to better address practical considerations of MPI. This includes the distribution of Fourier transforms and images across multiple nodes, a global fidelity constraint, and the explicit description of the MPI communication processes at different stages of the algorithms. Furthermore, we consider the pre-sampling of gridding kernels to reduce memory overhead significantly. These developments are absolutely critical to realise an efficient distributed MPI implemented that can scale to big-data in practice. While we plan to make improvements to performance in the future, this article describes details of the distributed Alternating Direction Method of Multipliers (ADMM), degridding, and wavelet transform algorithms with the public release of SOPT and PURIFY software versions, which makes these new distributed methods available to astronomers.

This manuscript starts by reviewing the interferometric imaging measurement equation in Section \ref{sec:ri_astro}. We then introduce sparse regularization for radio interferometric imaging in Section \ref{sec:sparse_reg}. In Section \ref{sec:prox_op} we introduce some basics of proximal calculus. In Section \ref{sec:ri_padmm} we introduce the serial Dual Forward-Backward based Alternating Direction Method of Multipliers (ADMM) algorithm. This sets the ground work for introducing computationally distributed wavelet and measurement operators and distributed ADMM algorithm in Section \ref{sec:dist_admm}, which have recently been made publicly available in the software packages PURIFY (\url{https://github.com/astro-informatics/purify}) and SOPT (\url{https://github.com/astro-informatics/sopt}). These algorithms make use of degridding and gridding, wavelet transforms, and proximal operators to reconstruct high quality images of the radio sky while communicating data between compute nodes of a computing cluster using MPI. We demonstrate the implementations of the distributed algorithms in PURIFY in Sections \ref{sec:benchmarks}, where we can distribute the data over a computing cluster. We use multi-threaded parallelization on a Graphics Processing Unit (GPU) or via OpenMP to parallelize across cores of a CPU node. Lastly, in Section \ref{sec:big_data} we show that pre-sampling the gridding kernel calculations can save memory and allow the application of ADMM to large data sets. We then show the application times for the distributed algorithms for up to 50 billion visibilities across the nodes of a computing cluster. We end with a conclusion in Section \ref{sec:conclusion}.}

\section{Radio Interferometric Measurement Equation}
\label{sec:ri_astro}
In this section we introduce concepts of radio interferometric imaging and the radio interferometric measurement equation. This measurement equation is calculated through the use of a linear measurement operator.

A radio interferometer is an array of radio antenna, where each pair of antenna results in a baseline and each baseline samples a Fourier coefficient of the brightness distribution across the radio sky. The separation of a baseline -- the baseline length -- determines the resolution scale of the Fourier coefficient. There are many measurement equations developed for radio interferometry \citep{mce08, car09, smi11, pri15} that are based on different approximations and different levels of accuracy. The van Cittert-Zernike theorem \citep{zer38} shows that for a small field of view and co-planar array, we find a simplified Fourier relation between visibilities and the sky brightness
\begin{equation}
	y(u, v) = \int^{1}_{-1} \int^{1}_{-1} x(l, m) a(l, m){\rm e}^{-2\pi i (lu + mv)}\,  {\rm d}l{\rm d}m\, ,
	\label{eq:meas_eq}
\end{equation}	
where the measurements $y$ are known as visibilities for the baseline separation $(u, v)$, $(l, m)$ are the sky coordinates, 
$x$ is the sky brightness, and operator $a$ includes direction dependent effects such as the primary beam and limits for the field of view. Equation \eqref{eq:meas_eq} consists of linear operations, such as integration, multiplication, and convolution. Any linear operation has a matrix representation after choosing a fixed basis and coordinate system, which is particularly useful when working with the discretized version of equation \eqref{eq:meas_eq}. 

To evaluate the measurement equation, and simulate the telescope, performing a Fourier transform is required. However, since $(u, v)$ do not lie on a regular grid, it is not possible to use the speed of the standard Fast Fourier Transform (FFT). Instead, a non-uniform FFT is used to evaluate the measurement equation, where an FFT is applied, and the visibilities are interpolated off the FFT grid \citep{fes03,tho08}. This process is traditionally known as degridding (or sometimes a prediction step). 

In a discrete setting, let $\bm{x} \in \mathbb{R}^{N}$ and $\bm{y} \in \mathbb{C}^{M}$ be the sky brightness and observed visibilities respectively.
A non-uniform FFT can be represented by the following linear operations 
\begin{equation}
	\bm{y} = \bm{\mathsf{W}}\bm{\mathsf{G}}\bm{\mathsf{F}}\bm{\mathsf{Z}}\bm{\mathsf{S}} \bm{x},
	\label{eq:matrix_equation}
\end{equation}
where $\bm{\mathsf{S}}$ represents a gridding correction and modeling of measurement independent instrumental effects, $\bm{\mathsf{Z}}$ represents zero padding of the image (to increase the resolution FFT, upsampling the Fourier grid), $\bm{\mathsf{F}}$ is an FFT, $\bm{\mathsf{G}}$ represents a convolution matrix that interpolates measurements off the grid, and $\bm{\mathsf{W}}$ are noise weights applied to the measurements. These linear operators represent the application of the measurement equation.
Let $\bm{\mathsf{\Phi}} =  \bm{\mathsf{W}}\bm{\mathsf{G}}\bm{\mathsf{F}}\bm{\mathsf{Z}}\bm{\mathsf{S}} \in \mathbb{C}^{M \times N}$, which is typically called the measurement operator for $N$ pixels and $M$ measurements. This measurement operator is described in detail in \citep{pra18}. 
Moreover, the linear measurement operator $\bm{\mathsf{\Phi}}$ has its adjoint operator $\bm{\mathsf{\Phi}}^\dagger$, which, in practice, consists of applying these operators in reverse (i.e. $\bm{\mathsf{S}}^\dagger\bm{\mathsf{Z}}^\dagger\bm{\mathsf{F}}^\dagger\bm{\mathsf{G}}^\dagger\bm{\mathsf{W}}^\dagger$). In this work, we assume that $\bm{y}$ are the weighted measurements, and the dirty map is defined as $\bm{\mathsf{\Phi}}^\dagger \bm{y}$.

\section{Sparse Regularization}
\label{sec:sparse_reg}
The previous section presents an inverse problem that has many solutions. Sparse regularization is a method that can estimate the radio sky brightness and isolate a single likely solution. In radio astronomy, the measurements have Gaussian uncertainty, leading to least squares minimization. To impose a penalty against over fitting of the radio sky, we can add a regularization term that penalizes models that over fit the measurements, i.e. a penalty that encourages the model to be sparse in parameters while fitting the radio sky. This leads to maximum a posteriori (MAP) estimation 
\begin{equation} \label{eq:MAP-estimate}
	\argmin_{\bm x} \gamma g(\bm{x}) + \|\bm{y} - \bm{\mathsf{\Phi}} \bm{x}\|_{\ell_2},
\end{equation}
where the least squares term is regularized by the function $g$ and $\gamma \geq 0$ determines the strength of the penalty against over fitting. The optimization problem presented with MAP estimation is known as the unconstrained problem.
One challenge of using MAP estimation to perform sparse regularization is choosing a proper regularization parameter $\gamma$ (although effective strategies do exist; \cite{per15}). 
The choice of $\gamma$, however, can be avoided after moving from the unconstrained problem in MAP estimation to the constrained problem
\begin{equation} \label{eq:model-con}
	\argmin_{\bm x} g(\bm{x}), \quad {\rm s.t.} \quad \|\bm{y} - \bm{\mathsf{\Phi}} \bm{x}\|_{\ell_2} \leq \epsilon,
\end{equation}
where $\epsilon$ is the error tolerance. One main advantage of the constrained objective function, compared to the unconstrained
form \eqref{eq:MAP-estimate}, is that the parameter $\epsilon$ can be estimated from $\bm{y}$ \cite{pra18}, and therefore could be easier to set than assign
a pertinent value for $\gamma$ in \eqref{eq:MAP-estimate}. Note, in practice, that the weights in $\bm{y}$ might be relative with no flux scale attached, 
or are not reliable, which will cause a difficulty for the constrained problem. On the other hand, progress is being made on methods that can 
estimate values of $\gamma$ for the unconstrained problem. 
It is also worth noticing that these two forms, \eqref{eq:MAP-estimate} and \eqref{eq:model-con}, have close relationship and, in some sense, are equivalent to each other
after assigning proper values for $\epsilon$ and $\gamma$.
The remainder of this work is focused on the constrained problem \eqref{eq:model-con} and we assume $\epsilon$ can be estimated.

\subsection{Analysis and Synthesis}
In the following we focus on using the $\ell_1$-norm for the function $g$ and require our solution to have positive real values, where the $\ell_p$-norm is defined by $\| \bm{x} \|_{\ell_p} = (\sum_i x_i^p)^{1/p}$ for $p > 0$. Additionally, we need to choose the representation of our signal to efficiently model the sky. This is done using a linear transform $\bm{\mathsf{\Psi}}$, with the convention that $\bm{x} = \bm{\mathsf{\Psi}} \bm{\alpha}$, where $\bm{\alpha}$ represents the coefficients of $\bm{x}$ under the basis or dictionary $\bm{\mathsf{\Psi}}$. A wavelet transform is convenient because it can efficiently represent structures as a function of scale and position. Moreover, $\bm{\mathsf{\Psi}}$ is not restricted to be a basis, but can be an over-complete frame containing a collection of transforms. In this work, we use a collection of wavelet transforms to model the radio sky, as done in \cite{car12,ono16,pra18,pra18a}.

The synthesis forms of the objective function for the unconstrained and constrained problems are respectively
\begin{equation} \label{eqn:model-syn-uncon}
	\bm{x}^* = \bm{\mathsf{\Psi}} \times \argmin_{\bm{\alpha}} \left\{ \|\bm{y} - \bm{\mathsf{\Phi}}\bm{\mathsf{\Psi}}\bm{\alpha}\|_{\ell_2}^2/2\sigma^2 + \gamma \| \bm{\alpha} \|_{\ell_1}, \quad {\rm s.t.} \quad\bm{\mathsf{\Psi}}\bm{\alpha} \in \mathbb{R}_+\right\}\, ,
\end{equation}
\begin{equation}
	\bm{x}^* = \bm{\mathsf{\Psi}} \times \argmin_{\bm{\alpha}} \left\{ \| \bm{\alpha} \|_{\ell_1}, \quad {\rm s.t.} \quad\|\bm{y} - \bm{\mathsf{\Phi}}\bm{\mathsf{\Psi}}\bm{\alpha}\|_{\ell_2} \leq \epsilon\quad \&  \quad\bm{\mathsf{\Psi}}\bm{\alpha} \in \mathbb{R}_+\right\}\, .
\end{equation}
The analysis forms of the objective function for the unconstrained and constrained problems are respectively
\begin{equation} 
	\bm{x}^* = \argmin_{\bm{x}} \left\{ \|\bm{y} - \bm{\mathsf{\Phi}}\bm{x}\|_{\ell_2}^2/2\sigma^2 + \gamma \| \bm{\mathsf{\Psi}}^\dagger\bm{x} \|_{\ell_1}, \quad {\rm s.t.} \quad\bm{x} \in \mathbb{R}_+\right\}\, ,
\end{equation}
\begin{equation}  \label{eqn:model-ana-con}
	\bm{x}^* = \argmin_{\bm{x}} \left\{  \| \bm{\mathsf{\Psi}}^\dagger\bm{x} \|_{\ell_1}, \quad {\rm s.t.}\quad\|\bm{y} - \bm{\mathsf{\Phi}}\bm{x}\|_{\ell_2} \leq \epsilon\quad \&  \quad\bm{x} \in \mathbb{R}_+ \right\}\, .
\end{equation}

In the synthesis form we solve for the wavelet coefficients $\bm{\alpha}$ directly and in the analysis form we solve for the pixel coefficients $\bm{x}$ directly.
In practice they provide different results depending on the problem to be solved \cite{ela06}. We follow the work of \cite{car12}, which uses an over-complete frame in the analysis setting and is typically found to provide better reconstruction quality than the synthesis setting. The objective function can be solved multiple-times after reweighting the $\ell_1$-norm in the analysis setting with an over-complete frame, using what is called Sparsity Averaging Reweighted Analysis (SARA) \cite{car12}.

Recent works have considered polarimetric \cite{aki17,bir18,bir18a} and spectral sparse image reconstruction \cite{abd16,deg16}. The works of \cite{bir18,bir18a} show that where polarimetric images are reconstructed as a four component vector of Stokes parameters $I$ (total intensity), $Q$ and $U$ (linear polarizations), and $V$ (circular polarization), it is possible to enforce the physical constraint that $I \geq \sqrt{Q^2 + U^2 + V^2}$. Such a constraint enforces physical structures on both total intensity and polarized intensity, increasing the physicality of the reconstructions. Additionally, it is possible to impose non-parametric structures on spectra, such as spectral smoothness or sparsity, increasing the fidelity across the spectrum.

The challenge in finding the global solution of these objective functions, \eqref{eqn:model-syn-uncon}--\eqref{eqn:model-ana-con}, is that they are non-differentiable (because of the non-differentiability of the $\ell_1$ regularization term) and are not always continuous (because they contain constraints). However, these objective functions have the property that they are convex and lower semi-continuous (l.s.c.). In the following sections, we introduce proximal operators, which provide tools and algorithms that can be used to find solutions to the above convex minimization problems.

\section{Proximal Operators}
\label{sec:prox_op}
In the previous section we introduced the convex objective functions  \eqref{eq:MAP-estimate} and \eqref{eq:model-con}, 
which need to be minimized to obtain a likely solution of the radio sky. 
When the problem is poised as minimization of a convex cost function, there are many convex optimization tools --
proximal operators and proximal algorithms among them -- on hand to solve it and find a global minimizer.
In the following, we briefly recall some concepts and operators of convex functions and convex sets, which are useful 
when discussing solutions to convex inverse problems. A more detailed introduction to these concepts can be found in \cite{boy04,com09,kom15}, and have been discussed in the context of radio interferometric imaging previously \cite{car14,car12,gir15,ono16,pra18,pra18a}. In this section, we review the basic mathematics of proximal operators, and introduce the closed-form solution of proximal operators used in this work.

Let $X$ be a vector space and $\Gamma_0(X)$ be the class of proper, l.s.c.\ convex functions that map from $X$ to $(-\infty, +\infty]$.
A function $h$ is convex when
\begin{equation}
	h(\alpha \bm{x}_1 + (1-\alpha)\bm{x}_2) \leq \alpha h(\bm{x}_1) + (1 - \alpha) h(\bm{x}_2), \quad \forall \bm{x}_1, \bm{x}_2 \in X, \forall \alpha \in [0 ,1],
\end{equation}
which is then true for $\forall h \in \Gamma_0(X)$.
The conjugate of $h \in \Gamma_0(X)$, denoted by $ h^* \in \Gamma_0(X)$, is defined as 
\begin{equation}
	h^*(\bm{m}) := \sup_{\bm{x} \in X} \left(\bm{m}^\top\bm{x} - h(\bm{x})\right).
\end{equation}
The conjugate can be used to map a convex objective function from the primal representation to its dual representation,
where both representations have the same optimal values when strong duality holds \cite{boy04,com09,boy15,kom15}.  
The subdifferential of $h$ at $\bm{x} \in X$, denoted by $\partial h(\bm{x})$, is defined as
\begin{equation} \label{def:subdiff}
	\partial h(\bm{x}) := \{\bm{u} \in X : h(\bm{z}) \ge h(\bm{x}) + \bm{u}^\top (\bm{z} - \bm{x}), \forall \bm{z} \in X  \}.
\end{equation}
When $h$ is differentiable, the subdifferential is a singleton containing the gradient $\nabla h$. If $\bm{0} \in \partial h(\bm{x})$ then $\bm{x}$ 
belongs to the set of global minimizers of $h$ \cite{com09}.

For $\forall h \in \Gamma_0(X)$ and any constant $\lambda >0$, the proximity operator of function $\lambda h$ at $\bm{v} \in X$, 
which is denoted by ${\rm prox}_{\lambda h}(\bm{v})$ and maps between $X \to X$, is defined as the solution of the minimization problem
\begin{equation}  	\label{eq:proximal}
	{\rm prox}_{\lambda h}(\bm{v}) = \argmin_{\bm{x}\in X}\left(\lambda h(\bm{x}) + \frac{1}{2}\|\bm{x} - \bm{v}\|^2_{\ell_2}\right).
\end{equation}
We see that ${\rm prox}_{\lambda h}(\bm{v})$  is a point that is chosen in $X$ by compromising between minimizing $h$ 
and being close to $\bm{v}$, where this compromise is weighted by $\lambda$. 
For large $\lambda$ more movement is taken towards minimizing $h$, and for small $\lambda$ less movement is taken from $\bm{v}$. 
The proximal operator in \eqref{eq:proximal} involves solving a minimization problem, which sometimes has a simple analytic form 
and sometimes not. When there is no analytic form it needs to be solved or estimated iteratively. It can be shown that the proximal operator is closely related 
to the subdifferential \eqref{def:subdiff}, being equivalent to the inverse operation $\left ( I  + \lambda\partial h \right)^{-1} (\bm{v})$ \cite{com09}. 

When applied to a convex function, the proximal operator can be used to find a global minimizer through the recursive iteration. 
This is because the proximal operator is what is known as firmly non-expansive. More importantly it is a contraction, meaning repeated application of the proximal operator
\begin{equation}
	\bm{x}^{k+1} = {\rm prox}_{\lambda h}(\bm{x}^k)
\end{equation}
will converge to a fixed point that minimizes $\lambda h$ and therefore also 
minimizes $h$; that is, $\bm{x} = {\rm prox}_{\lambda h}(\bm{x})$ if and only if $\bm{x}$ minimizes $h$ \cite{boy04,com09}. 

The proximal operator has plenty of useful properties. For example, the proximal operator for the translation, the semi-orthogonal linear transform 
and the convex conjugation are
\begin{equation} \label{eqn:prox-trans}
{\rm prox}_{\lambda h(\cdot+\bm{a})}(\bm{x}) = {\rm prox}_{\lambda h}(\bm{x} + \bm{a}) - \bm{a}, \quad \forall \bm{a} \in X,
\end{equation}
\begin{equation}
	{\rm prox}_{\lambda h(\bm{\mathsf{L}}(\cdot))}(\bm{x}) = \bm{x} + \bm{\mathsf{L}}^\dagger \left ({\rm prox}_{\lambda h}(\bm{\mathsf{L}}\bm{x})  -\bm{\mathsf{L}}\bm{x} \right), 
		\quad \bm{\mathsf{L}} \bm{\mathsf{L}}^\dagger = \bm{\mathsf{I}}
\end{equation}
and
\begin{equation}
\label{eq:Moreau_decomposition}
	{\rm prox}_{\lambda h^*}(\bm{x}) = \bm{x} - \lambda {\rm prox}_{\lambda^{-1}h}(\bm{x}/\lambda ),
\end{equation}
respectively. The property for convex conjugation is also known as Moreau decomposition. Refer to \cite{boy04,com09} and references therein for other properties and more details. Typically, it is difficult to obtain a closed form of the proximal operator for two functions $f + g$. The algorithms in the following section split the algorithm into solving for $f + g$ given the proximal operator of $f$ and $g$ separately, and are typically called proximal splitting algorithms. First, we introduce closed forms of proximal operators that are used in radio interferometric imaging (but more examples are listed in \cite{boy04,com09}).

In this work, we focus on $\ell_1$ regularized least squares, i.e., using the $\ell_1$ prior for $g$ in the constrained problem \eqref{eq:model-con}. 
We need to minimize an $\ell_1$-norm with the condition that the solution lies within an $\ell_2$-ball with the size of our error $\epsilon$, while being real or positive valued. This can be mathematically stated as
\begin{equation}
  	\bm{x}^\star = \argmin_{\bm{x}} \left\{ \|\bm{\mathsf{\Psi}}^\dagger\bm{x}\|_{\ell_1} + \iota_{\mc{C}} (\bm{x}) + \iota_{\mathcal{B}^\epsilon_{\ell_2}(\bm{y})} (\bm{\Phi}\bm{x}) \right\}\, ,
	\label{l1-constrained}
\end{equation}
where we normally take $\mc{C} = \mathbb{R}_+^N$, and $\ell_2$-ball $\mathcal{B}^\epsilon_{\ell_2}$ to be the closed ball of radius $\epsilon$, and $\iota_\mc{C}(\bm{x})$ is the indicator function for $\bm{x}$ being in $\mc{C}$ which will be detailed below. We now present the proximal operators needed to minimize this objective function.

\subsection{Indicator Function}
Fix any nonempty closed convex set $\mc{C}$, on which we define the indicator function as
\begin{equation} 	\label{eq:infinite step function}
	\iota_{\mc{C}}(\bm{x}) :=
	\begin{cases}
		0,        & \bm{x} \in \mc{C},    \\
		+\infty,        & \bm{x} \notin \mc{C}. 
	\end{cases} 
\end{equation}
We recall the projection operator $\mc{P}_{\mc{C}}$, i.e.
\begin{equation}
	\mc{P}_{\mc{C}} (\bm{x}) := \argmin_{\bm{v} \in \mc{C}} \| \bm{x} - \bm{v} \|_{\ell_2}^2.
\end{equation}
If $\mc{C} \subseteq X$, then we have $\iota_{\mc{C}} \in \Gamma_0(X)$ and
{
\begin{equation}
	\mc{P}_{\mc{C}} (\bm{x}) = \argmin_{\bm{v} \in \mc{C}} \| \bm{x} - \bm{v} \|_{\ell_2}^2 
		=   \argmin_{\bm{v} \in X} \left\{\iota_{\mc{C}} (\bm{v}) + \| \bm{x} - \bm{v} \|_{\ell_2}^2 \right\}
		= {\rm prox}_{\iota_{\mc{C}}}(\bm{x}).
\end{equation}
}
Therefore, the proximal operator can be regarded as an extension of the projection operator \cite{com09}. 
The indicator function is useful for e.g. restricting a cost function to a set of solutions, or enforcing real or positive values on the solutions
as assumptions for an image of the radio sky. 

\subsection{Fidelity Constraint}
Let the closed $\ell_2$-ball $\mathcal{B}^\epsilon_{\ell_2}$ centered at $\bm{z} \in X$ with radius $\epsilon$ be the set
\begin{equation}
	\mathcal{B}^\epsilon_{\ell_2}(\bm{z}) := \left\{\bm{v} \in X : \| \bm{z} - \bm{v} \|_{\ell_2} \leq \epsilon \right\}\, .
\end{equation}
Then the proximal operator of an $\ell_2$-ball centered at zero reads
\begin{align} \label{eqn:prox-l2-ball}
\begin{split}
	{\rm prox}_{\mathcal{B}^\epsilon_{\ell_2}(0)}(\bm{x})
		& = \argmin_{\bm{v} \in X} \left\{ \iota_{\mathcal{B}^\epsilon_{\ell_2}(0)}(\bm{v}) +  \frac{1}{2}\| \bm{v} - \bm{x} \|_{\ell_2}^2 \right\} \\
		& =
		\begin{cases}
		\bm{x},        & \bm{x} \in \mathcal{B}^\epsilon_{\ell_2}(0),   \\
		\frac{\bm{x}}{\|\bm{x}\|} \epsilon,      & \bm{x} \notin \mathcal{B}^\epsilon_{\ell_2}(0).
		\end{cases}
\end{split}		
\end{align}
In detail, when $\bm{x} \in \mathcal{B}^\epsilon_{\ell_2}(0)$, we have ${\rm prox}_{\mathcal{B}^\epsilon_{\ell_2}(0)}(\bm{x}) = \bm{x}$ straightforwardly; 
when $\bm{x} \notin \mathcal{B}^\epsilon_{\ell_2}(0)$, computing ${\rm prox}_{\mathcal{B}^\epsilon_{\ell_2}(0)}(\bm{x})$ 
is to find a $\bm{v} \in \mathcal{B}^\epsilon_{\ell_2}(\bm{0})$ such that it minimizes $\| \bm{v} - \bm{x} \|_{\ell_2}^2$. From the triangle inequality, we require that $\bm{v}$ is parallel to $\bm{x}$ for it to be a minimizer.
It follows that we can scale $\bm{x}$ into $ \mathcal{B}^\epsilon_{\ell_2}(0)$ to obtain the explicit representation of ${\rm prox}_{\mathcal{B}^\epsilon_{\ell_2}(0)}(\bm{x})$
shown in \eqref{eqn:prox-l2-ball}. Using the translation property of the proximal operator in \eqref{eqn:prox-trans}, we can find the proximal operator of 
an $\ell_2$-ball centered at $\bm{z}$, i.e.,
\begin{equation} 	\label{eq:l2ball_prox}
	\proj_{\mc{B}}^{\epsilon}(\bm{z}) := 
	{\rm prox}_{\mathcal{B}^\epsilon_{\ell_2}(\bm{z})}(\bm{x}) {=} 
	\begin{cases}
		\bm{x},       & \bm{x} - \bm{z} \in \mathcal{B}^\epsilon_{\ell_2}(\bm{0}),   \\
		\frac{\bm{x} - \bm{z}}{\|\bm{x} - \bm{z} \|} \epsilon + \bm{z},   & \bm{x} - \bm{z} \not\in \mathcal{B}^\epsilon_{\ell_2}(\bm{0}).
	\end{cases}
\end{equation}

\subsection{Promoting Sparsity}
The $\ell_1$-norm is the sum of the absolute values of all components of a vector.
Since it is convex and can promote sparsity when serving as a prior distribution or regularization, it is widely used in 
signal/image processing and has been shown highly effective in radio astronomy.   

The proximal operator of the $\ell_1$-norm reads
\begin{align}
\begin{split}
	{\rm prox}_{\lambda\| \cdot \|_{\ell_1}}(\bm{x}) & = \argmin_{\bm{v} \in X} \left\{ \lambda \| \bm{v} \|_{\ell_1} + \frac{1}{2}\| \bm{v} - \bm{x} \|_{\ell_2}^2 \right\} \\
		& = \soft_{\lambda}(\bm{x}).
\end{split}		
\end{align}
Here $\soft_{\lambda}(\bm{x})$ is the soft thresholding of vector $\bm{x} = (x_1, \cdots, x_i, \cdots)$, which is defined as 
\begin{equation}
	\soft_{\lambda}(\bm{x}) = (\soft_{\lambda}(x_1), \cdots, \soft_{\lambda}(x_i), \cdots),
\end{equation}
where
\begin{equation} 	
	\soft_{\lambda}(x_i) {=} 
	\begin{cases}
		0,       & |x_i| \le \lambda,   \\
		\frac{x_i (|x_i| - \lambda)}{|x_i|},   &  |x_i| > \lambda.
	\end{cases}
\end{equation}
An intuitive derivation can be found by differentiating $\lambda | v | + \frac{1}{2}( v - x )^2$ with respect to $v$, and then solving for 
\begin{equation} 	
	x {=} 
	\begin{cases}
		0,       & v = 0,   \\
		v + \lambda   &  v > 0, \\
		v - \lambda   &  v < 0.
	\end{cases}
\end{equation}
Then case by case we find $x = 0$ when $v = 0$, $x > \lambda$ when $v > 0$ and $ x < -\lambda$ when $v < 0$. Rearranging the formula for each case, we arrive at soft thresholding as the proximal operator for the $\ell_1$-norm.

\subsection{Summary}
This section has provided an introduction to proximal operators and examples of their closed-form solutions that are commonly used for interferometric imaging of real observations \cite{ono16,pra18,dab18}. Proximal operators are especially powerful when the objective function is non-smooth, which is often required to enforce physicality on the solution. One important example in polarimetric imaging, but not detailed here,
is where epigraphical projection techniques are used to construct a proximal operator that will project onto the set of solutions that contains $I \geq \sqrt{U^2 + Q^2 + V^2}$ \cite{bir18,bir18a}.

We have provided proximal operators for a function $f$, but we often need to minimize an addition of functions, e.g. $f + g$. In the next section, we show how to solve for the minimizer of $f + g$ when the proximal operators of $f$ and $g$ are known separately.

\section{Sparse Regularization using Dual Forward-Backward ADMM}
\label{sec:ri_padmm}
As mentioned in \eqref{l1-constrained}, the standard constrained radio interferometry solution with $\ell_1$ (sparse) regularization can be stated as
\begin{equation}
  	\bm{x}^\star = \argmin_{\bm{x}} \left\{ \|\bm{\mathsf{\Psi}}^\dagger\bm{x}\|_{\ell_1} + \iota_{\mc{C}} (\bm{x}) + \iota_{\mathcal{B}^\epsilon_{\ell_2}(\bm{y})} (\bm{\Phi}\bm{x}) \right\} \, ,
	\label{global-min-problem-admm}
\end{equation}
with $\mathcal{B}^\epsilon_{\ell_2}(\bm{y}) = \{ \bm{z} \in \bb{C}^{M}: \| \bm{z} - \bm{y} \|_{\ell_2} \leq \epsilon \}$ being the set that satisfies the fidelity constraint and 
$\mc{C} = \mathbb{R}_+^N$  is the set that represents the positive and real constraint.

Let $\bm{r}$ be the slack variable with the constraint $\bm{r} = \bm{\Phi}\bm{x}$. As described in Section \ref{sec-admm}, to solve the above problem \eqref{global-min-problem-admm}, ADMM can be applied
by minimizing the Lagrangian of problem \eqref{global-min-problem-admm} corresponding to $\bm{x}$ and $\bm{r}$ alternatively, i.e.,
\begin{equation}
	\min_{\bm{x}} \left\{ \mu \left[\|\bm{\mathsf{\Psi}}^\dagger\bm{x}\|_{\ell_1} + \iota_{\mc{C}} (\bm{x})\right] + \frac{1}{2} \big\|\bm{\Phi} \bm{x} - (\bm{r} - \bm{s}) \big\|_{\ell_2}^2 \right\},
	\label{admm-basic-min-steps-x}
\end{equation}
\begin{equation}
	\min_{\bm{r}} \left\{ \mu \left[\iota_{\mathcal{B}^\epsilon_{\ell_2}(\bm{y})}(\bm{r})\right] + \frac{1}{2} \big\| \bm{r} - (\bm{\Phi} \bm{x} + \bm{s}) \big\|_{\ell_2}^2 \right\} ,
	\label{admm-basic-min-steps-r}
\end{equation}
where $\bm{s}$ represents the Lagrangian multiplier.
Algorithm \ref{alg-admm} shows the Dual Forward-Backward ADMM algorithm used to solve problem \eqref{global-min-problem-admm}. 
Recall that it is the same as the standard ADMM algorithm, but uses Dual Forward-Backward splitting with a Forward-Backward step to minimize the subproblem \eqref{admm-basic-min-steps-x}. The distributed version of this algorithm is presented in \cite{ono16}. The serial version of this algorithm has been implemented in PURIFY and applied in \cite{pra18} to simulated and real observations from radio interferometric telescopes previously.

\begin{algorithm}[t]
	\caption{Dual Forward-Backward \ac{admm}.\newline
	The Dual Forward-Backward ADMM algorithm without MPI implementation.
	Lines 3--4 evaluate the $\ell_2$-ball proximal operator (constraining the solution to the $\ell_2$-ball), which is to address the solution of the subproblem \eqref{admm-basic-min-steps-r}. Line 5 is the Lagrangian dual variable update, connecting the two minimization problems \eqref{admm-basic-min-steps-x} and \eqref{admm-basic-min-steps-r}. Lines 6--7 are a Forward-Backward step, which is to address the solution of the subproblem \eqref{admm-basic-min-steps-x}; particularly, line 6 is the forward (gradient) step, and line 7 is the backward step which is solved using the Dual Forward-Backward algorithm, as described between lines 9--16.
	}
	\label{alg-admm}
																																																																		  
	\begin{algorithmic}[1]
		\small
		\Given{$\bm{x}^{(0)}, \bm{r}^{(0)}, \bm{s}^{(0)}, \bm{q}^{(0)}, \gamma, \rho, \varrho$}
		\RepeatFor{$t=1,\ldots$}
		\State $\ds \bm{v}^{(t)} = \bm{\mathsf{\Phi}}\bm{x}^{(t-1)}$
		\State $\bm{r}^{(t)} = \proj_{\mc{B}}^\epsilon \Big( \bm{v}^{(t)} + \bm{s}^{(t-1)} \Big)$
		\State $\bm{s}^{(t)} = \bm{s}^{(t-1)} + \varrho \big(\bm{v}^{(t)} - \bm{r}^{(t)} \big)$
		\State $\ds \tilde{\bm{x}}^{(t)} = \bm{x}^{(t-1)} - \rho \bm{\mathsf{\Phi}}^\dagger\bigg(\bm{v}^{(t)} - \bm{r}^{(t)} + \bm{s}^{(t)} \bigg)$
		\State $\bm{x}^{(t)} =$ \FuncCall{DualFB}{$\tilde{\bm{x}}^{(t)}, \gamma$}
		\Until {\bf convergence \normalfont}
		\vspace{5px}
		\Func{DualFB}{$\bm{z}, \gamma$}
		\Given{$\bm{d}^{(0)}_j, \eta$}
		\State $\ds \bar{\bm{z}}^{(0)} = \proj_{\mc{C}}  \big( \bm{z} \big)$
		\RepeatFor{$k=1,\ldots$}
		\State $\bm{d}^{(k)} = \frac{1}{\eta} \Bigg(\! \identity - \soft_{\gamma} \!\!\Bigg)\! \Big( \eta \bm{d}^{(k-1)} + \bm{\mathsf{\Psi}}^\dagger \bar{\bm{z}}^{(k-1)} \Big)$
		\vspace{-5px}
		\State $\ds \bar{\bm{z}}^{(k)} = \proj_{\mc{C}}  \bigg(\bm{z}^{(k - 1)} -  \bm{\mathsf{\Psi}} \bm{d}^{(k)} \ec \bigg)$
		\Until {\bf convergence \normalfont}
		\EndFunc{$\bar{\bm{z}}^{(k)}$}
	\end{algorithmic}
\end{algorithm}

\section{Distributed Dual Forward-Backward ADMM}
\label{sec:dist_admm}
In the previous sections, we covered serial proximal optimization algorithms and serial operators. It is well known that these algorithms can be distributed (see \cite{boy11,kom15,ono16} and references therein).

In the remainder of this work, we describe the details for how to modify these algorithms to be distributed over a computing cluster using the MPI standard commonly known as MPI. For clarity, we describe MPI implementations of operators in PURIFY and SOPT. The measurements and MPI processes are distributed across the nodes of a computing cluster.

\subsection{MPI Framework}
The MPI standard is a framework where multiple process of the same program are run concurrently, communicating data and variables at sync points. This is commonly referred to as distributed memory parallelism. There are many independent processes (nodes) with their own data, but they can send messages containing data between them. This is different from the more typical shared memory parallelism, where a single process has access to all the data, but executes multiple threads for sections of the program (such as a loop with independent iterations). However, hybrids of shared and distributed memory parallelism are not uncommon, where nodes on a computing cluster send messages while performing multi-threaded operations. Please see \cite{sni98} for a formal reference on MPI\footnote{Official versions of the MPI standard can be found online at \url{https://www.mpi-forum.org/docs/}.}.

The MPI framework contains a total number of process $n_d$, each with a rank $0 \leq j < n_d$, all connected by a communicator for sending and receiving data. The most basic methods of a communicator consist of send and receive operations between individual processes. However, typically sending and receiving is performed in collective send and receive operations:

{\bf Broadcast (one to many) --}
Send a copy of a variable (scalar or array) from the root node to all nodes. 

{\bf Scatter (one to many) --}
Scatter is where a root process contains an array; different sections of this array are sent to different nodes. The root process does not keep the sent data.

{\bf Gather (many to one) --} Gather is where the root process receives data from all nodes. This could be sections of an array, or variables that are combined into an array on the root process.

{\bf All to All (many to many) --}
All to all is where data is communicated between all nodes at once. Each process sends and receives. This could be single variables or sections of arrays.

{\bf Reduce (many to one) --}
Reduce, or performing a reduction, is where a binary operation (assumed to be associative and commutative) is efficiently performed with a variable or array over the cluster with the result sent to the root process. Summation of variables across nodes is a common example of this. However, logical operations and max/min operations are also common.

{\bf All reduce (many to many) --}
All reduce is equivalent to a reduction, but the result is broadcasted to all nodes from the root process. All reduce with summation is called an all sum all operation.

The operation to broadcast a copy of $\bm{x}$ onto each node can be represented by the linear operation
\begin{equation}
		\begin{bmatrix}
		\bm{\mathsf{I}}_1\\
		\vdots \\
		\bm{\mathsf{I}}_{n_{\rm d}}
	\end{bmatrix}\bm{x}
\end{equation}
where $\bm{\mathsf{I}}_j$ is an $N\times N$ identity matrix. The adjoint of this operation is a reduction
\begin{equation}
	\bm{x}_{\rm sum} = \begin{bmatrix}
		\bm{\mathsf{I}}_1 & \dots &\bm{\mathsf{I}}_{n_{\rm d}}
	\end{bmatrix}	\begin{bmatrix}
		\bm{x}_1\\
		\vdots \\
		\bm{x}_{n_{\rm d}}
	\end{bmatrix}\, .
\end{equation}
It is possible to view other MPI operations of sending data between nodes in the context of linear mappings. 

\subsection{Distributed Visibilities}
\label{sec:dist_data_op}
The visibilities can be loaded on a root process then sorted into groups that are scattered to each node. This process splits and sorts the measurement vector $\bm{y}$ into groups $\bm{y}_j$, where $j$ is the rank of a process:
\begin{equation}
	\bm{y} = \begin{bmatrix}
		\bm{y}_1\\
		\vdots \\
		\bm{y}_{n_{\rm d}}
	\end{bmatrix}\, .
\end{equation}
In this work, we sort the visibilities $\bm{y}$ via ordering them by baseline length and dividing $\bm{y}$ into sections of equal size $\bm{y}_j$ to be scattered to each node. However, it is also possible to have each MPI process to read a different set of measurements. In principle, the weights and $uvw$ coordinates are scattered with the visibilities.

If there is too much data to load the {measurements} onto one node, the data can be loaded in sections and then scattered to each node. After the data has been distributed, sorting into groups can be done using logical reductions, and then distributed to each node using an all to all operation. This has been done with the $w$-stacking algorithm in \cite{pra18a}.

\subsection{Distributed Measurement Operator}
\label{sec:dist_measure_op}
For each group of visibilities $\bm{y}_j$ on node $j$, there is a corresponding measurement operator $\bm{\mathsf{\Phi}}_j$. However, there are many ways to relate $\bm{\mathsf{\Phi}}_j$ to the measurement operator for $\bm{y}$, $\bm{\mathsf{\Phi}}$; we show two examples.

\subsubsection{Distributed Images}
\label{sec:mo_dist_image}
We can relate the MPI measurement operator to the serial operators by
\begin{equation}
\bm{\mathsf{\Phi}} =
	\begin{bmatrix}\bm{\mathsf{\Phi}}_1 & &\\
		 & \ddots & \\
		 & & \bm{\mathsf{\Phi}}_{n_{\rm d}}
	 \end{bmatrix}\begin{bmatrix}
		\bm{\mathsf{I}}_1\\
		\vdots \\
		\bm{\mathsf{I}}_{n_{\rm d}}
	\end{bmatrix}\, .
\end{equation}

The forward operator can be expressed simply as independent measurement operators applied in parallel after broadcasting $\bm{x}$:
\begin{equation}
	\begin{bmatrix}
		\bm{y}_1\\
		\vdots \\
		\bm{y}_{n_{\rm d}}
	\end{bmatrix}
	=\begin{bmatrix}\bm{\mathsf{\Phi}}_1 & &\\
		 & \ddots & \\
		 & & \bm{\mathsf{\Phi}}_{n_{\rm d}}
	 \end{bmatrix}\begin{bmatrix}
		\bm{\mathsf{I}}_1\\
		\vdots \\
		\bm{\mathsf{I}}_{n_{\rm d}}
	\end{bmatrix}
	\bm{x}\, .
\end{equation}
The adjoint operator can be expressed as the adjoint of independent measurement operators applied in parallel, followed by a reduction
\begin{equation}
		\bm{x}_{\rm dirty} =	\begin{bmatrix}
		\bm{\mathsf{I}}_1 & \dots &\bm{\mathsf{I}}_{n_{\rm d}}
	\end{bmatrix}\begin{bmatrix}\bm{\mathsf{\Phi}}_1^\dagger & &\\
		 & \ddots & \\
		 & & \bm{\mathsf{\Phi}}_{n_{\rm d}}^\dagger
	 \end{bmatrix}
		\begin{bmatrix}
		\bm{y}_1\\
		\vdots \\
		\bm{y}_{n_{\rm d}}
	\end{bmatrix}\, .
\end{equation}
However, with the MPI framework, it is efficient to always have a copy of the same image on each node so that other image domain operations can be performed in parallel (i.e. wavelet transforms). This can be ensured by combining the broadcast and reduction in a single all sum all operation during the adjoint.
We work with the forward operator that applies each measurement operator independently on each node, with a copy of $\bm{x}$ located on each node
\begin{equation}
	\begin{bmatrix}
		\bm{y}_1\\
		\vdots \\
		\bm{y}_{n_{\rm d}}
	\end{bmatrix}
	=\begin{bmatrix}\bm{\mathsf{\Phi}}_1 & &\\
		 & \ddots & \\
		 & & \bm{\mathsf{\Phi}}_{n_{\rm d}}
	 \end{bmatrix}\begin{bmatrix}
		\bm{x}\\
		\vdots \\
		\bm{x}
	\end{bmatrix}\, ,
\end{equation}
and the adjoint operation can be performed by applying the adjoint of each measurement operator independently followed by an all sum all
\begin{equation}
		\begin{bmatrix}
		\bm{x}_{\rm dirty}\\
		\vdots \\
		\bm{x}_{\rm dirty}
	\end{bmatrix}=	\begin{bmatrix}
		\bm{\mathsf{I}}_1\\
		\vdots \\
		\bm{\mathsf{I}}_{n_{\rm d}}
	\end{bmatrix}\begin{bmatrix}
		\bm{\mathsf{I}}_1 & \dots &\bm{\mathsf{I}}_{n_{\rm d}}
	\end{bmatrix}\begin{bmatrix}\bm{\mathsf{\Phi}}_1^\dagger & &\\
		 & \ddots & \\
		 & & \bm{\mathsf{\Phi}}_{n_{\rm d}}^\dagger
	 \end{bmatrix}
		\begin{bmatrix}
		\bm{y}_1\\
		\vdots \\
		\bm{y}_{n_{\rm d}}
	\end{bmatrix}\, .
\end{equation}
We can normalize the operator with the operator norm, by using the power method to estimate the largest eigenvalue, and remove arbitrary scaling due to $n_{\rm d}$ and other normalization factors. {This implementation is unique to this work and is different from the method discussed next which is used in \cite{ono16}.}

\subsubsection{Distributed FFT Grid Sections}
\label{sec:mo_dist_grid}
Another method {we implemented}, which is discussed in \cite{ono16}, is to distribute the grid points of the FFT grid, where the degridding can be performed on each node. This can be performed using a scatter and gather operation from a root process. We can define the operation of distributing the necessary grid points using the operators $\bm{\mathsf{M}}_j \in \mathbb{R}^{B_j\times 2N}$, where $B_j$ is the number of non zero columns of $\bm{\mathsf{G}}_j$. Additionally, we can remove the zero columns of $\bm{\mathsf{G}}_j$, such that $\bm{\mathsf{G}}_j \in \mathbb{R}^{M_j \times B_j}$.

The measurement operator is defined by
\begin{equation}
	\begin{bmatrix}
		\bm{y}_1\\
		\vdots \\
		\bm{y}_{n_{\rm d}}
	\end{bmatrix}
	=\begin{bmatrix}\bm{\mathsf{W}}_1\bm{\mathsf{G}}_1 & &\\
		 & \ddots & \\
		 & & \bm{\mathsf{W}}_{n_{\rm d}}\bm{\mathsf{G}}_{n_{\rm d}}
	 \end{bmatrix}\begin{bmatrix}
		\bm{\mathsf{M}}_1\\
		\vdots \\
		\bm{\mathsf{M}}_{n_{\rm d}}
	\end{bmatrix}\bm{\mathsf{F}}\bm{\mathsf{Z}}\bm{\mathsf{S}}
	\bm{x}\, .
\end{equation}
$[\bm{\mathsf{M}}_1^\top, \cdots, \bm{\mathsf{M}}_{n_{\rm d}}^\top]^\top$	
	can be seen as scattering the FFT grid points from the root process to the other nodes.
The adjoint can be seen as gathering and summing gridded FFT grid points to the root process. While this method appears to reduce communication, this has the disadvantage that the result of the adjoint ends up only on the root process. In practice, this means a broadcast is eventually required after the adjoint of this measurement operator so that further image domain operations can be performed in parallel.

\subsection{Distributed Wavelet Operator}
\label{sec:dist_wavelet_op}
The MPI wavelet operator can be distributed for each wavelet basis in the dictionary. Using the convention that $\bm{x} = \bm{\mathsf{\Psi}}\bm{\alpha}$, each wavelet representation can be arranged as
\begin{equation}
	\bm{\alpha} = \begin{bmatrix} 
\bm{\alpha}_1 \\
\vdots \\
\bm{\alpha}_{n_{\rm w}}
	\end{bmatrix}\, ,
\end{equation}
for $n_{\rm w}$ wavelet transforms. 
From this definition, it follows that each inverse transform is performed independently with a reduction at the end
\begin{equation}
\bm{\mathsf{\Psi}} =
\begin{bmatrix}
		\bm{\mathsf{I}}_1 &
		\dots &
		\bm{\mathsf{I}}_{n_{\rm w}}
	\end{bmatrix}
	\begin{bmatrix}\bm{\mathsf{\Psi}}_1 & &\\
		 & \ddots & \\
		 & & \bm{\mathsf{\Psi}}_{n_{\rm w}}
	 \end{bmatrix}\, .
\end{equation}
However, like with the distributed image measurement operator, we combine the reduction and broadcasting as an all sum all. In practice, we use the forward operation
\begin{equation}
\begin{bmatrix}
		\bm{x} \\
		\vdots \\
		\bm{x}
	\end{bmatrix} =\begin{bmatrix}
		\bm{\mathsf{I}}_1\\
		\vdots \\
		\bm{\mathsf{I}}_{n_{\rm w}}
	\end{bmatrix}
\begin{bmatrix}
		\bm{\mathsf{I}}_1 &
		\dots &
		\bm{\mathsf{I}}_{n_{\rm w}}
	\end{bmatrix}
	\begin{bmatrix}\bm{\mathsf{\Psi}}_1 & &\\
		 & \ddots & \\
		 & & \bm{\mathsf{\Psi}}_{n_{\rm w}}
	 \end{bmatrix}\begin{bmatrix}
		\bm{\alpha}_1\\
		\vdots \\
		\bm{\alpha}_{n_{\rm w}}
	\end{bmatrix}\, .
\end{equation}
The adjoint operation is
\begin{equation}
\begin{bmatrix}
		\bm{\alpha}_1\\
		\vdots \\
		\bm{\alpha}_{n_{\rm d}}
	\end{bmatrix} =
	\begin{bmatrix}\bm{\mathsf{\Psi}}_1^\dagger & &\\
		 & \ddots & \\
		 & & \bm{\mathsf{\Psi}}_{n_{\rm w}}^\dagger
	 \end{bmatrix}\begin{bmatrix}
		\bm{x}\\
		\vdots \\
		\bm{x}
	\end{bmatrix}\, .
\end{equation}

\subsection{Distributed Proximal Operator}
\label{sec:dist_proximal_op}
The proximal operators for the $\ell_1$-norm, $\ell_2$-ball, and convergence criteria may require communication between nodes, which is discussed in this section.
\subsubsection{Sparsity and Positivity Constraint}
The $\ell_1$-proximal norm does not need a communicator in itself. However, $\bm{\mathsf{\Psi}}$ contains more than one wavelet transform. The proximal operator for the $\ell_1$-norm is solved iteratively using the Dual Forward-Backward method. The objective function that proximal operator minimizes can be computed to check that the iterations have converged. For a given $\bm{x}$, the proximal operator returns
{
\begin{equation}
	\argmin_{\bm{z}}\left[\iota_{\mathbb{R}_+^N}(\bm{z}) + \|\bm{\mathsf{\Psi}}^\dagger \bm{z}\|_{\ell_1} + \frac{1}{2\gamma}\| \bm{x} - \bm{z} \|_{\ell_2}\right]\, .
\end{equation}
}
To assert that the Dual Forward-Backward method has converged to a minimum when calculating the proximal operator requires checking the variation of the $\ell_1$-norm; calculating the $\ell_1$-norm requires an MPI all sum all operation over wavelet coefficients. Another assertion that can be made is that the relative variation of $\bm{x}$ is close to zero, which requires no communication.

\subsubsection{Fidelity Constraint}
In the constrained minimization problem, the solution is constrained to be within the $\ell_2$-ball through the proximal operator ${\rm prox}_{\mathcal{B}^\epsilon_{\ell_2}(\bm{y})}(\bm{v})$. However, this proximal operator requires calculating the $\ell_2$-norm of the residuals $\|\bm{v} - \bm{y}\|_{\ell_2}$. When the visibilities are distributed on each node $\bm{y}_i$, this calculation requires an all sum all.

However, if each node constrains the solution to an independent local $\ell_2$-ball {using} ${\rm prox}_{\mathcal{B}^{\epsilon_j}_{\ell_2}(\bm{y}_j)}(\bm{v}_j)$ {with} radius $\epsilon_j$, where $\epsilon = \sqrt{\sum_{j = 1}^{n_d} \epsilon_j^2}$. This solution will also lie within the global $\ell_2$-ball {where we have used} ${\rm prox}_{\mathcal{B}^\epsilon_{\ell_2}(\bm{y})}(\bm{v})$, which can be shown using the triangle inequality. This requires less communication (introducing a new $\epsilon_j$ for each node is suggested \cite{ono16}). However, the communication overhead is negligible because much larger data rates are being communicated. Furthermore, using the global $\ell_2$-ball is more robust in convergence rate as it is independent of how the measurements are grouped across the nodes.

\subsection{Distributed Convergence}
\label{sec:dist_convergence}
There are multiple methods that can be used to check that the solution has converged. For example, when the relative difference of the solution between iterations is small, i.e. $\|\bm{x}^{(i)} - \bm{x}^{(i - 1)}\|_{\ell_2}/\|\bm{x}^{(i)}\|_{\ell_2} < \delta$ for a small $\delta$; when the relative difference of the objective function between iterations is small; and the condition that the residuals of the solution lie within the $\ell_2$-ball\footnote{A feature of ADMM is that it will not ensure that the residuals lie in the $\ell_2$-ball for each iteration but it will converge to this condition.}. These convergence criteria need to be communicated across the nodes. The convergence criteria need to be chosen carefully, since the quality of the output image can be degraded if the iterations have not converged sufficiently.

\subsection{Distributed ADMM}
With PURIFY, we build on the previous sections and combine the MPI distributed linear operators and proximal operators to solve the radio interferometric imaging inverse problem. 
The previous section discusses how to distribute $\bm{\mathsf{\Phi}}$ for distributed $\bm{y}_j$, and how to distribute $\bm{\mathsf{\Psi}}$ for distributed wavelet coefficients. In Algorithms \ref{alg-admm-image} and \ref{alg-admm-grid}, we outline MPI algorithms that use two variations of the measurement operator. Algorithm \ref{alg-admm-image} uses an all sum all in the adjoint of the measurement operator following Section \ref{sec:mo_dist_image} and Algorithm \ref{alg-admm-grid} performs an FFT on the root node distributing parts of the grid following Section \ref{sec:mo_dist_grid}. In practice we recommend using Algorithm \ref{alg-admm-image} as it can be easily modified to efficiently model wide-field effects, as demonstrated in \cite{pra18a}. Furthermore, Algorithm \ref{alg-admm-image} is simpler to implement.
\begin{algorithm}[t]
	\caption{Distributed Image (Dual Forward-Backward \ac{admm}):\newline
		Every node has access to a global $\ell_2$-ball proximal and a serial version of the measurement operator $\bm{\Phi}_j$. After the adjoint of the measurement operator is applied, an $\bm{\rm AllSumAll}$ is performed over the returned image of each node $j$, then each node has the combined image. An $\bm{\rm AllSumAll}$ is also used after the forward wavelet operator $\bm{\Psi}_j$. Communication is needed in calculation of $\proj_{\mc{B}}^\epsilon$ with an $\bm{\rm AllSumAll}$ in the $\ell_2$-norm of the residuals. Using instead $\proj_{\mc{B}_j}^{\epsilon_j}$ removes this communication overhead but changes the minimization problem.
	}
	\label{alg-admm-image}
																																																																		  
	\begin{algorithmic}[1]
		\small
		\Given{$\bm{x}^{(0)}, \bm{r}^{(0)}_j, \bm{s}^{(0)}_j, \bm{q}_j^{(0)}, \gamma, \rho, \varrho$}
		\RepeatFor{$t=1,\ldots$}
		\ParFor{$\forall j \in \{1, \ldots, n_{\rm{d}}\}$}
		\State $\ds \bm{v}^{(t)}_j = \bm{\mathsf{\Phi}}_j\bm{x}^{(t-1)}$
		\State $\bm{r}^{(t)}_j = \proj_{\mc{B}}^\epsilon \Big( \bm{v}^{(t)}_j + \bm{s}^{(t-1)}_j \Big)$
		\State $\bm{s}^{(t)}_j = \bm{s}^{(t-1)}_j + \varrho \big(\bm{v}^{(t)}_j - \bm{r}^{(t)}_j \big)$
		\State $\bm{q}^{(t)}_j = \bm{\mathsf{\Phi}}_j^\dagger\bigg(\bm{v}^{(t)}_j - \bm{r}^{(t)}_j + \bm{s}^{(t)}_j \bigg)$
		\State $\ds \tilde{\bm{x}}^{(t)} = \bm{x}^{(t-1)} - \rho \bm{{\rm AllSumAll}_j}(\bm{q}^{(t)}_j)$
		\State $\bm{x}^{(t)} =$ \FuncCall{DualFB}{$\tilde{\bm{x}}^{(t)}, \gamma$}
		\EndParFor
		\Until {\bf convergence \normalfont}
		\vspace{5px}
		\Func{DualFB}{$\bm{z}, \gamma$}
		\Given{$\bm{d}^{(0)}_j, \eta$}
		\State $\ds \bar{\bm{z}}^{(0)} = \proj_{\mc{C}}  \big( \bm{z} \big)$
		\RepeatFor{$k=1,\ldots$}
		\ParFor{$\forall j \in \{1, \ldots, n_{\rm{w}}\}$}
		\State $\bm{d}^{(k)}_j = \frac{1}{\eta} \Bigg(\! \identity - \soft_{\gamma} \!\!\Bigg)\! \Big( \eta \bm{d}^{(k-1)}_j + \bm{\mathsf{\Psi}}_j^\dagger \bar{\bm{z}}^{(k-1)} \Big)$
		\vspace{-5px}
		\State $\ds \bar{\bm{z}}^{(k)} = \proj_{\mc{C}}  \bigg(\bm{z} -  \bm{{\rm AllSumAll}_j}\left(\bm{\mathsf{\Psi}}_j \bm{d}^{(k)}_j \right) \ec \bigg)$
		\EndParFor
		\Until {\bf convergence \normalfont}
		\EndFunc{$\bar{\bm{z}}^{(k)}$}
	\end{algorithmic}
\end{algorithm}

\begin{algorithm}[ht]
	\caption{Distributed Fourier Grid (Dual Forward-Backward \ac{admm}):\newline
		Every node has access to a global $\ell_2$-ball proximal operator. The measurement operator is split. First, the root process computes $\bm{\mathsf{F}}\bm{\mathsf{Z}}\bm{\mathsf{S}}$, and scatters parts of the FFT grid $\bm{b}_j^{(t)}$ to each node. Each node then applies $\bm{\mathsf{G}}_j$ to predict the visibilities for the $j^{\rm th}$ node. After the $\ell_2$-ball proximal operator is applied, each node applies $\bm{\mathsf{G}}_j^\dagger$, then the root node gathers and adds the result. Then an update image is broadcast to the other nodes, which is needed for \FuncCall{DualFB}{$\tilde{\bm{x}}^{(t)}, \gamma$}. The rest of the algorithm is as in Algorithm \ref{alg-admm-image}.
	}
	\label{alg-admm-grid}
																																																																		  
	\begin{algorithmic}[1]
		\small
		\Given{$\bm{x}^{(0)}, \bm{r}_j^{(0)}, \bm{s}_j^{(0)}, \bm{q}_j^{(0)}, \gamma, \rho, \varrho$}
		\RepeatFor{$t=1,\ldots$}																																						 
		\ParFor{$\forall j \in \{1, \ldots, n_{\rm{d}}\}$}
		\State {\bf Root process only}: $\ds \tilde{\bm{b}}^{(t)} = \bm{\mathsf{F}}\bm{\mathsf{Z}}\bm{\mathsf{S}} \bm{x}^{(t-1)}$
		\State $\ds \bm{b}_j^{(t)} = \bm{{\rm Scatter}_j}(\bm{\mathsf{M}}_j \tilde{\bm{b}}^{(t)})$
		\State $\bm{v}^{(t)}_j = \bm{\mathsf{G}}_j \bm{b}^{(t)}_j$
		\State $\bm{r}_j^{(t)} = \proj_{\mc{B}}^{\epsilon} \Big(\bm{v}^{(t)}_j  + \bm{s}_j^{(t-1)} \Big)$
		\State $\bm{s}_j^{(t)} = \bm{s}_j^{(t-1)} + \varrho \big( \bm{v}^{(t)}_j - \bm{r}_j^{(t)} \big)$
		\State $\bm{q}_j^{(t)} = \bm{\mathsf{G}}^\dagger_j \bigg( \bm{v}^{(t)}_j - \bm{r}_j^{(t)} + \bm{s}_j^{(t)} \bigg)$ 
		\State {\bf Root process only}: $\bm{q}_j^{(t)} = \bm{{\rm Gather}_j}(\bm{q}_j^{(t)})$
		\State {\bf Root process only}: $\ds \tilde{\bm{x}}^{(t)} = \bm{x}^{(t-1)} - \rho \bm{\mathsf{Z}}^\dagger \bm{\mathsf{F}}^\dagger \sum_{j=1}^{n_{\rm{d}}} \bm{\mathsf{M}}_j^\dagger \bm{q}_j^{(t)}$
		\State $\tilde{\bm{x}}^{(t)} = \bm{\rm Broadcast}(\tilde{\bm{x}}^{(t)})$
		\State $\bm{x}^{(t)} =$ \FuncCall{DualFB}{$\tilde{\bm{x}}^{(t)}, \gamma$}
		\EndParFor
		\Until {\bf convergence \normalfont}
		\vspace{5px}
		\Func{DualFB}{$\bm{z}, \gamma$}
		\Given{$\bm{d}^{(0)}_i, \eta$}
		\State $\ds \bar{\bm{z}}^{(0)} = \proj_{\mc{C}}  \big( \bm{z} \big)$
		\RepeatFor{$k=1,\ldots$}
		\ParFor{$\forall j \in \{1, \ldots, n_{\rm{w}}\}$}
		\State $\bm{d}^{(k)}_j = \frac{1}{\eta} \Bigg(\! \identity - \soft_{\gamma} \!\!\Bigg)\! \Big( \eta \bm{d}^{(k-1)}_j + \bm{\mathsf{\Psi}}_j^\dagger \bar{\bm{z}}^{(k-1)} \Big)$
		\vspace{-5px}
		\State $\ds \bar{\bm{z}}^{(k)} = \proj_{\mc{C}}  \bigg(\bm{z} -  \bm{{\rm AllSumAll}_j}\left(\bm{\mathsf{\Psi}}_j \bm{d}^{(k)}_j \right) \ec \bigg)$
		\EndParFor
		\Until {\bf convergence \normalfont}
		\EndFunc{$\bar{\bm{z}}^{(k)}$}
	\end{algorithmic}
\end{algorithm}
\subsection{Global Fidelity Constraint ADMM}
When the measurements are spread across the various nodes, communication is required to ensure that the same $\ell_2$-ball constraint is enforced across all measurements. The proximal operator for the $\ell_2$-ball is
\begin{equation}
	\proj_{\mc{B}}^{\epsilon}(\bm{z}_j)  \overset{\Delta}{=} \left\{ 
	\begin{array}{cl}
		\epsilon \frac{\bm{z}_j - \bm{y}_j}{\sqrt{\bm{{\rm AllSumAll}_j}(\|\bm{z}_j - \bm{y}_j\|_{\ell_2}^2)}} + \bm{y_j} & \qquad \sqrt{\bm{{\rm AllSumAll}_j}(\|\bm{z}_j - \bm{y}_j\|_{\ell_2}^2)} > \epsilon    \\
		\bm{z}_j \qquad \qquad                                                    & \qquad \sqrt{\bm{{\rm AllSumAll}_j}(\|\bm{z}_j - \bm{y}_j\|_{\ell_2}^2)} \leq \epsilon 
	\end{array}\right.,
	\label{proj-L2-global}
\end{equation}
{The global $\epsilon$ with a measurement operator is different from other work, and requires a communication step not described in other work.}
\subsection{Local Fidelity Constraint ADMM}
\label{sec:local-ell2}
We can split the single $\ell_2$-ball into many, and restate a new constrained problem, i.e.,
\begin{equation}
  	\bm{x}^\star = \argmin_{\bm{x}} \left\{  \|\bm{\mathsf{\Psi}}^\dagger\bm{x}\|_{\ell_1} + \iota_{\mc{C}} (\bm{x}) + \sum_{j = 1}^{n_{\rm d}}\iota_{\mc{B}_j} (\bm{\Phi}_j\bm{x}) \right\}\, .
	\label{local-min-problem-admm}
\end{equation}
In particular, the alternating minimization involving the slack variable $\bm{r}$ is split into solving each $\bm{r}_j$ independently
\begin{equation}
	\min_{\bm{r}_j} \left\{  \mu \left[\iota_{\mc{B}_j}(\bm{r}_j)\right] + \frac{1}{2} \big\| \bm{r}_j - \bm{\Phi}_j \bm{x} - \bm{s}_j \big\|_{\ell_2}^2 \right\} .
	\label{split-admm-basic-min-steps-r}
\end{equation}
Each $\ell_2$-ball proximal operator acts on a different section of $\bm{y}_j$, so they can be performed in parallel with no communication \cite{ono16}
\begin{equation}
	\proj_{\mc{B}_j}^{\epsilon_j}(\bm{z}_j)  \overset{\Delta}{=} \left\{ 
	\begin{array}{cl}
		\epsilon_j \frac{\bm{z}_j - \bm{y}_j}{\|\bm{z}_j - \bm{y}_j\|_{\ell_2}} + \bm{y}_j & \qquad \|\bm{z}_j - \bm{y}_j\|_{\ell_2} > \epsilon_j    \\
		\bm{z}_j \qquad \qquad                                                    & \qquad \|\bm{z}_j - \bm{y}_j\|_{\ell_2} \leq \epsilon_j 
	\end{array}\right .
	\label{proj-L2-local}.
\end{equation}
By replacing $\proj_{\mc{B}}^{\epsilon}(\bm{z}_j)$ with $\proj_{\mc{B}_j}^{\epsilon_j}(\bm{z}_j)$ in Algorithms \ref{alg-admm-image} and \ref{alg-admm-grid}, the communication needed can be reduced.

The reduced communication overhead due to the local $\ell_2$-ball constraint is negligible compared to the overhead of the $\bm{\rm Broadcast}$ and $\bm{\rm AllSumAll}$ operations performed on $\bm{x}$, since $n_{\rm d} \ll N$. Additionally, there is a drawback when the convergence is sensitive to the distribution of $\bm{y}_j$, which is not the case for the global $\ell_2$-ball ADMM. We thus advocate using the global fidelity constraint.

\section{Algorithm Performance using PURIFY}
\label{sec:benchmarks}
We have implemented the MPI ADMM algorithms from the previous sections in PURIFY and SOPT. In this section, we benchmark the performance against the non-distributed counterpart \cite{pra18}, to show that such methods can decrease the time required for each iteration. We also implement and benchmark a GPU implementation of the measurement operator against its CPU implementation, to show that GPU implementations can further increase the performance (which can be used in conjunction with the MPI algorithms).

\subsection{PURIFY Software Package}
PURIFY has been developed as a software package that will perform distributed sparse image reconstruction of radio interferometric observations to reconstruct a model of the radio sky. The sparse convex optimization algorithms and MPI operations have been implemented in a standalone library known as SOPT. Previous versions of PURIFY are described in \cite{car14,pra18}. In this section, we describe the latest release of {PURIFY (Version 3.1.0)} \cite{pra19} and latest release of {SOPT (Version 3.1.0)} \cite{pra19b} that accompany this article. You can download and find the latest details on PURIFY at \url{https://github.com/astro-informatics/purify} and SOPT at \url{https://github.com/astro-informatics/sopt}.

PURIFY and SOPT have been developed using the C++11 standard. We use the software package Eigen for linear algebra operations \cite{eigen}. OpenMP is used to increase performance of the FFT, discrete planar wavelet transforms, and sparse matrix multiplication. The separable 2d discrete Daubechies wavelet transforms (1 to 30) have been implemented using a lifting scheme (more details on wavelet transforms can be found in \cite{dau98,mal99}), and have been multi-threaded over application of filters. The sparse matrix multiplication is multi-threaded over rows, requiring the sparse matrix to be stored in row-major order for best performance. To perform operations on a GPU, we use the library ArrayFire, which can be used with a CPU, OpenCL, or CUDA back-end \cite{Yalamanchili2015}. Within SOPT, we have implemented various MPI functionality (all sum all, broadcast, gather, scatter, etc.) to interface with data types and communicate the algorithm operations across the cluster. It is possible to read the measurements (and associated data) using UVFITS or CASA Measurement Set (MS) formats. The UVFITS format follows the format set by \cite{gre16}. The output images are saved as two dimensional FITS file images, as a sine projection. Currently, the distributed algorithm supported is ADMM. Furthermore, $w$-projection and $w$-stacking algorithms are supported for wide-fields of view and are described in \cite{pra18a}.

\subsection{Distribution of Visibilities}
PURIFY can read visibilities $\{\bm{y}_i\}_{i = 1}^{n_{\rm d}}$, and scatter them to each node on the cluster. How these measurements are distributed is not important when using the global $\ell_2$-ball constraint. However, when using local $\ell_2$-ball constraints, the way the visibilities are grouped for each node could make a difference to convergence, where it could be better to keep similar baselines on the same node. We do this by assigning different nodes to different regions of the FFT grid, or by assigning different nodes to regions in baseline length $\sqrt{u^2 + v^2}$. However, when using the $w$-stacking algorithm $k$-means with MPI is used to redistribute the visibilities over the cluster using an all to all operation, as discussed in \cite{pra18a}.

\subsection{Benchmark Timings}
In the remainder of this section, we time the operations of the MPI algorithms. We use Google Benchmark\footnote{\url{https://github.com/google/benchmark}} to perform the timings of the mean and standard deviation for each operation benchmarked. The times provided are in real time (incorporating communication), not CPU time, since multi-threaded operations are sensitive to this difference. Each benchmark configuration was timed for 10 runs, providing a mean and standard deviation used for timings and errors in the sections that follow.

The computing cluster Legion at University College London was used to measure the benchmark timings. We used the Type U nodes on Legion, which work as a 16 core device with 64GB RAM (160 Dell C6220 nodes - dual processor, eight cores per processor\footnote{More details can be found at \url{https://wiki.rc.ucl.ac.uk/wiki/RC_Systems\#Legion_technical_specs}.}).

In the benchmarks, the root node generates a random Gaussian density sampling distribution of baselines $(u, v)$, ranging from $\pm \pi$ along each axis. The weights $\bm{\mathsf{W}}_j$ and baseline coordinates $(\bm{u}_j, \bm{v}_j)$ are distributed to nodes $1 \leq j \leq n_{\rm d}$. This allows the construction of $\bm{\mathsf{W}}_j\bm{\mathsf{G}_j}$ on each node. We use the Kaiser-Bessel kernel as the interpolation (anti-aliasing) convolution kernel for $\bm{\mathsf{G}_j}$, with a minimum support size of $J = 4$ (see \cite{pra18} for more details). The identical construction of $\bm{\mathsf{F}}\bm{\mathsf{Z}}\bm{\mathsf{S}}$ can then be performed on each node or the root node (depending on the algorithm), and allow us to apply $\bm{\mathsf{\Phi}}$ in each of the MPI algorithms.

\subsection{MPI Measurement Operator Benchmarks}
The $\bm{\rm AllSumAll}(\bm{x})$ and $\bm{\rm Broadcast}(\bm{x})$ in the $\bm{\mathsf{\Phi}}^\dagger$ operations will be expensive in communication overheads for large $N$. Additionally, the calculation of the FFT $\bm{\mathsf{F}}$ does not take advantage of MPI and will have the cost $\mathcal{O}(N\log N)$, albeit the FFT is multi threaded using FFTW and OpenMP to provide performance improvements. It is more likely that the time taken to compute the FFT will take longer than the communication of the image at large $N$. If we evenly distribute the visibilities so that each node has $M_j = M/n_{\rm d}$, the computational complexity of the sparse matrix multiplication $\bm{\mathsf{G}}_j$ reduces to $\mathcal{O}(M_jJ^2)$ per node, providing a large advantage at large $M$ and $n_{\rm d}$. 

\begin{figure}
	\center
	\includegraphics[width=0.5\textwidth]{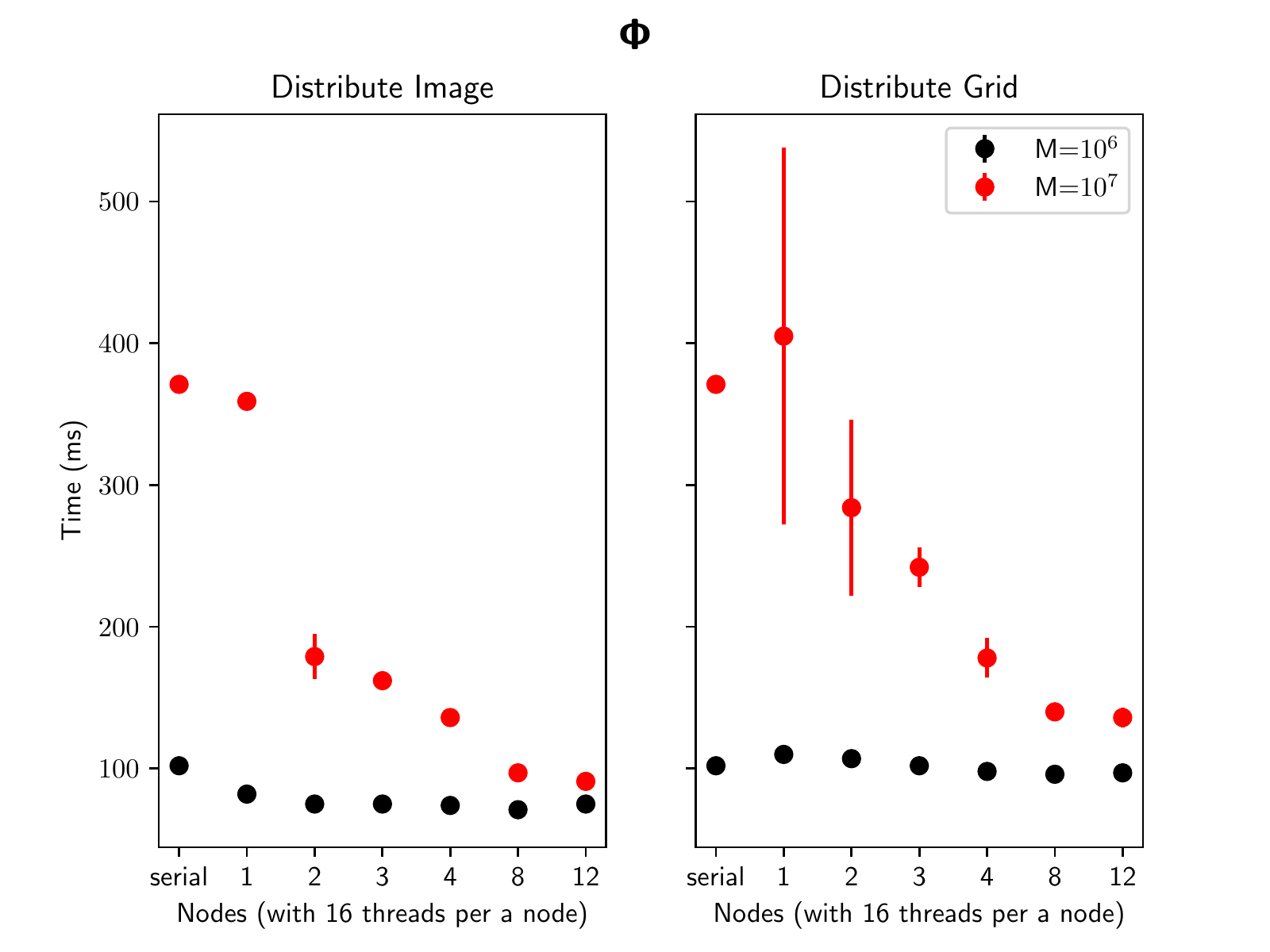}\includegraphics[width=0.5\textwidth]{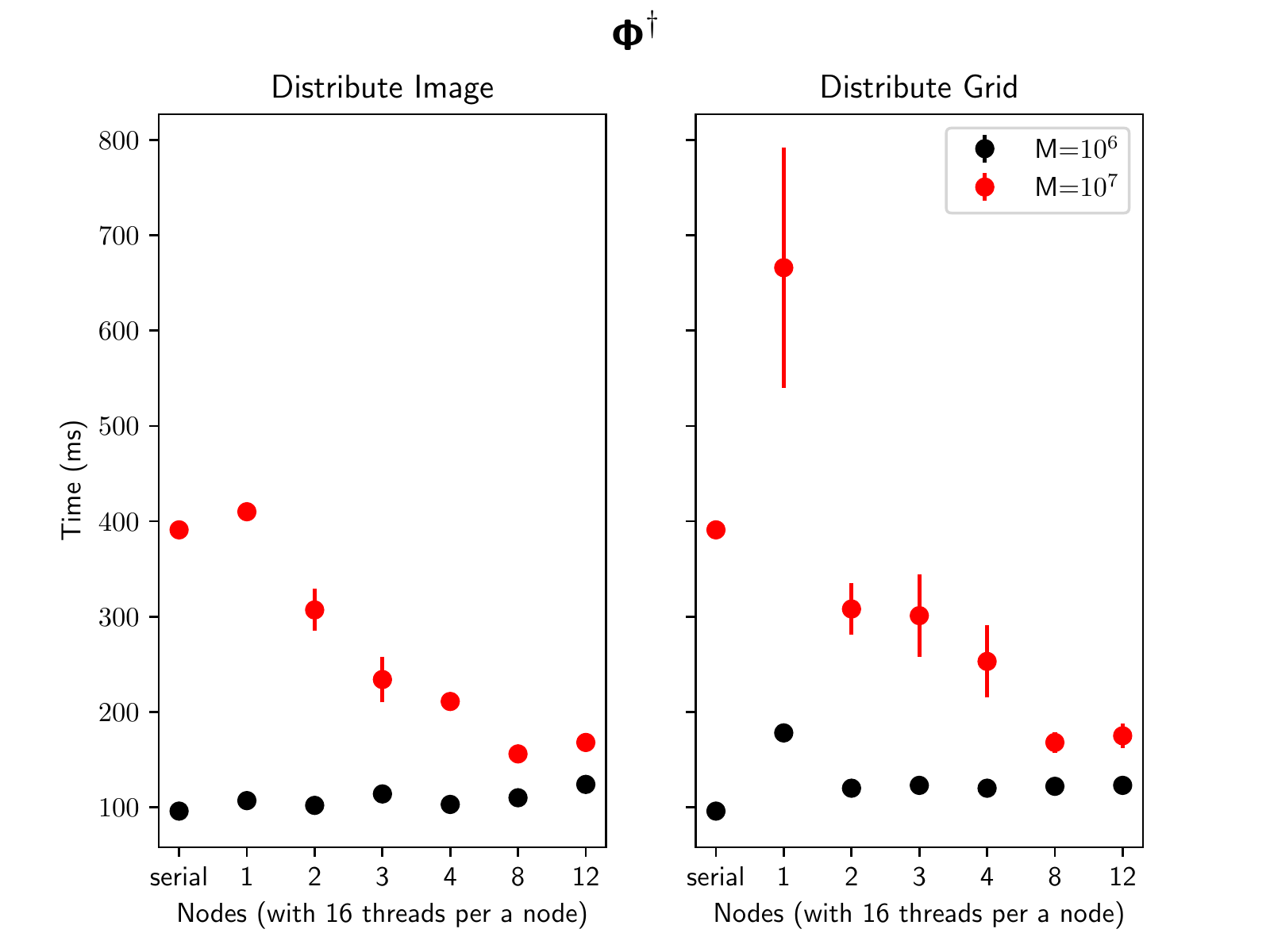}
	\caption{Time to apply forward $\bm{\mathsf{\Phi}}$ (left two plots) and adjoint $\bm{\mathsf{\Phi}}^\dagger$ (right two plots) as a function of the number of MPI nodes, benchmarked against the non MPI (serial) implementation. We fix the number of visibilities and image size at $N = 1024\times 1024$, $M \in \{10^6, 10^7\}$. On the left the MPI implementation corresponds to using an all sum all MPI operation in the adjoint described in Section \ref{sec:mo_dist_image}; on the left MPI implementation corresponds to distribution of the grid from the root node, as described in Section \ref{sec:mo_dist_grid}.}
	\label{fig:mo_bench}
\end{figure}

We benchmark the MPI $\bm{\mathsf{\Phi}}$ and $\bm{\mathsf{\Phi}}^\dagger$ implementations against the non-distributed equivalent using PURIFY. We use $10^6$ and $10^7$ visibilities, and a fixed image size of $N = 1024\times 1024$. We vary the number of nodes from $1, 2, 3, 4, 8, 12$. Results are shown in Figure \ref{fig:mo_bench}. For $10^6$ visibilities there is no improvement on the measurement operator performance for each MPI implementation. However, for $10^7$ it is clear that increasing the number of nodes increases the performance. The saturation for $n_{\rm d} \geq 5$ can be explained by the computational cost of the FFT $\bm{\mathsf{F}}$ being greater than the sparse matrix multiplication $\bm{\mathsf{G}}_j$. For small numbers of nodes $n_{\rm d}$, i.e. 1 or 2, the application time is less reliable, but for larger $n_{\rm d}$ it becomes more stable. We also find that distributing sections of the grid (described in Section \ref{sec:mo_dist_grid}) is more expensive at low $n_{\rm d}$ than distributing the image (described in Section \ref{sec:mo_dist_image}).

\subsection{MPI Wavelet Operator Benchmarks}
\begin{figure}
	\center
	\includegraphics[width=0.5\textwidth]{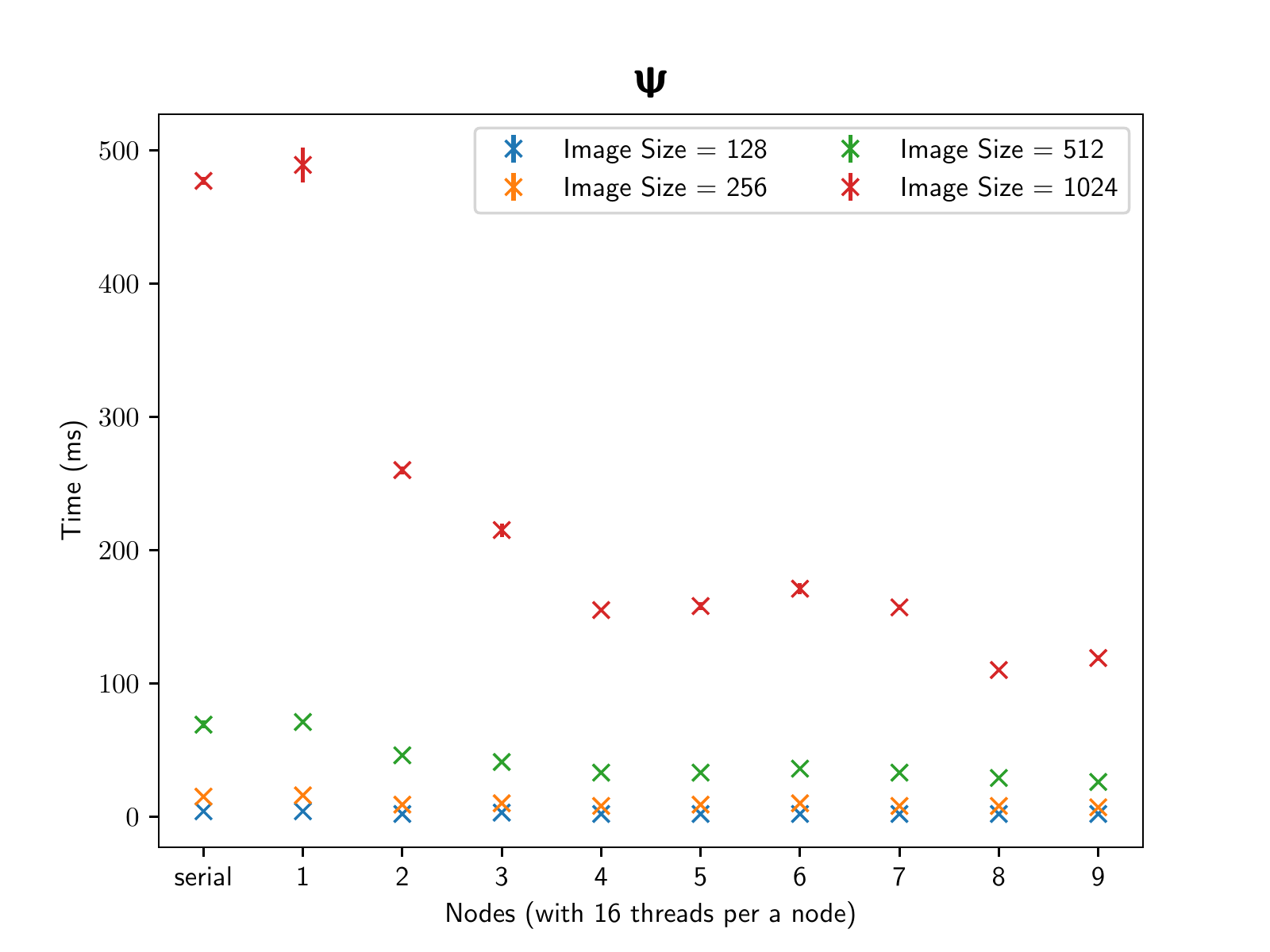}\includegraphics[width=0.5\textwidth]{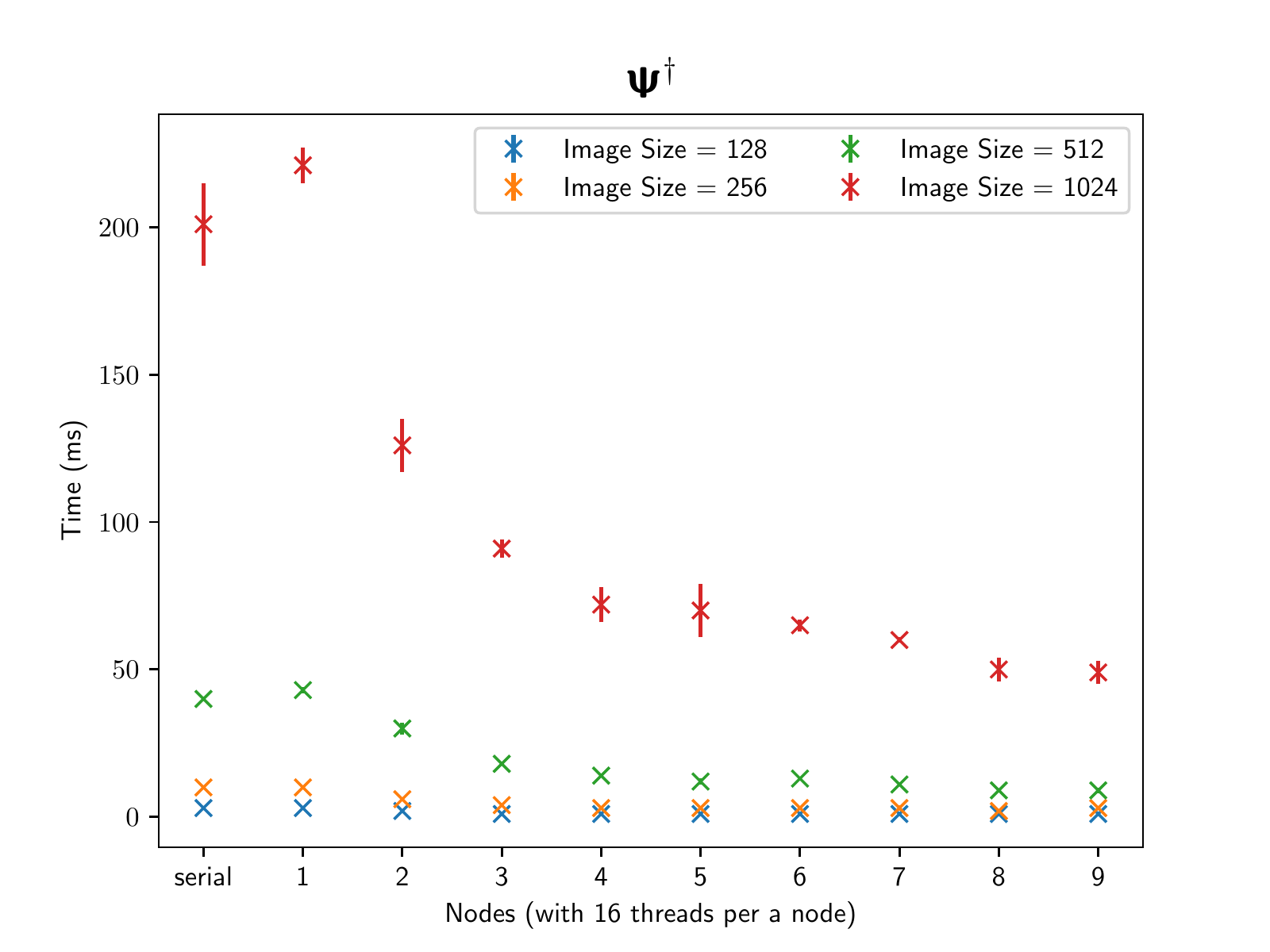}
	\caption{Time to apply forward $\bm{\mathsf{\Psi}}$ (left) and adjoint $\bm{\mathsf{\Psi}}^\dagger$ (right) as a function of the number of MPI nodes, benchmarked against the non MPI (serial) implementation. The forward operator requires 2 times more calculations than the adjoint due to the up sampling operations. Distributing the wavelet transforms across the nodes greatly decreases the time for calculation.}
	\label{fig:wlo_bench}
\end{figure}

Like the measurement operator $\bm{\mathsf{\Phi}}$, the wavelet operator $\bm{\mathsf{\Psi}}$ requires an $\bm{\rm AllSumAll}(\bm{x})$ operation. However, even with multi-threaded operations in the wavelet transform, computing $\bm{\mathsf{\Psi}}_j$ is time consuming. When $n_{\rm w} > n_{\rm d}$, multiple wavelet transforms are performed on some of the nodes, for example when $n_{\rm w} = 2n_{\rm d}$ there are two wavelet transforms per node. In many cases we expect that the numbers of nodes is greater than the number of wavelet transforms, i.e. $n_{\rm d} \geq n_{\rm w}$, and the maximum benefit from MPI distribution of wavelet transforms can be seen. This can be seen in Figure \ref{fig:wlo_bench}, where there is a performance improvement with using MPI to distribute the wavelet transforms across nodes.

In the benchmarks, we use $n_{\rm w} = 9$ where $\bm{\mathsf{\Psi}}_0$ is a Dirac basis and $\bm{\mathsf{\Psi}}_1$ to $\bm{\mathsf{\Psi}}_8$ are 2d (the product of 1d) Db wavelets 1 to 8. We perform the wavelet transform to three levels. Increasing the number of wavelet levels requires more computation, but much of this computation is in the first few levels. Furthermore, the low pass and high pass filters in the Db increase with size from 1 to 8, meaning Db 8 requires more computation than Db 7 at each wavelet level (but we have found the time difference small). 
The forward operator $\bm{\mathsf{\Psi}}$ requires up-sampling, meaning that it requires a factor of 2 times more computation than the adjoint $\bm{\mathsf{\Psi}}^\dagger$. 

\subsection{MPI Algorithm Benchmarks}
As a demonstration the impact of the MPI operators, we benchmark the Algorithms \ref{alg-admm-image} and Algorithm \ref{alg-admm-grid} against the serial Algorithm \ref{alg-admm} (equivalent to $n_{\rm d} = 1$). We fix the number of visibilities and image size at $N = 1024 \times 1024$, $M \in \{10^6,10^7\}$.

We use local $\ell_2$-ball constraints for each node as described in Section \ref{sec:local-ell2}. However, PURIFY also provides the ability to use the global $\ell_2$-constraint. In practice, we do not find much difference in computation time between using a local or global $\ell_2$-constraint.

\begin{figure}
	\center
	\includegraphics[width=0.5\textwidth]{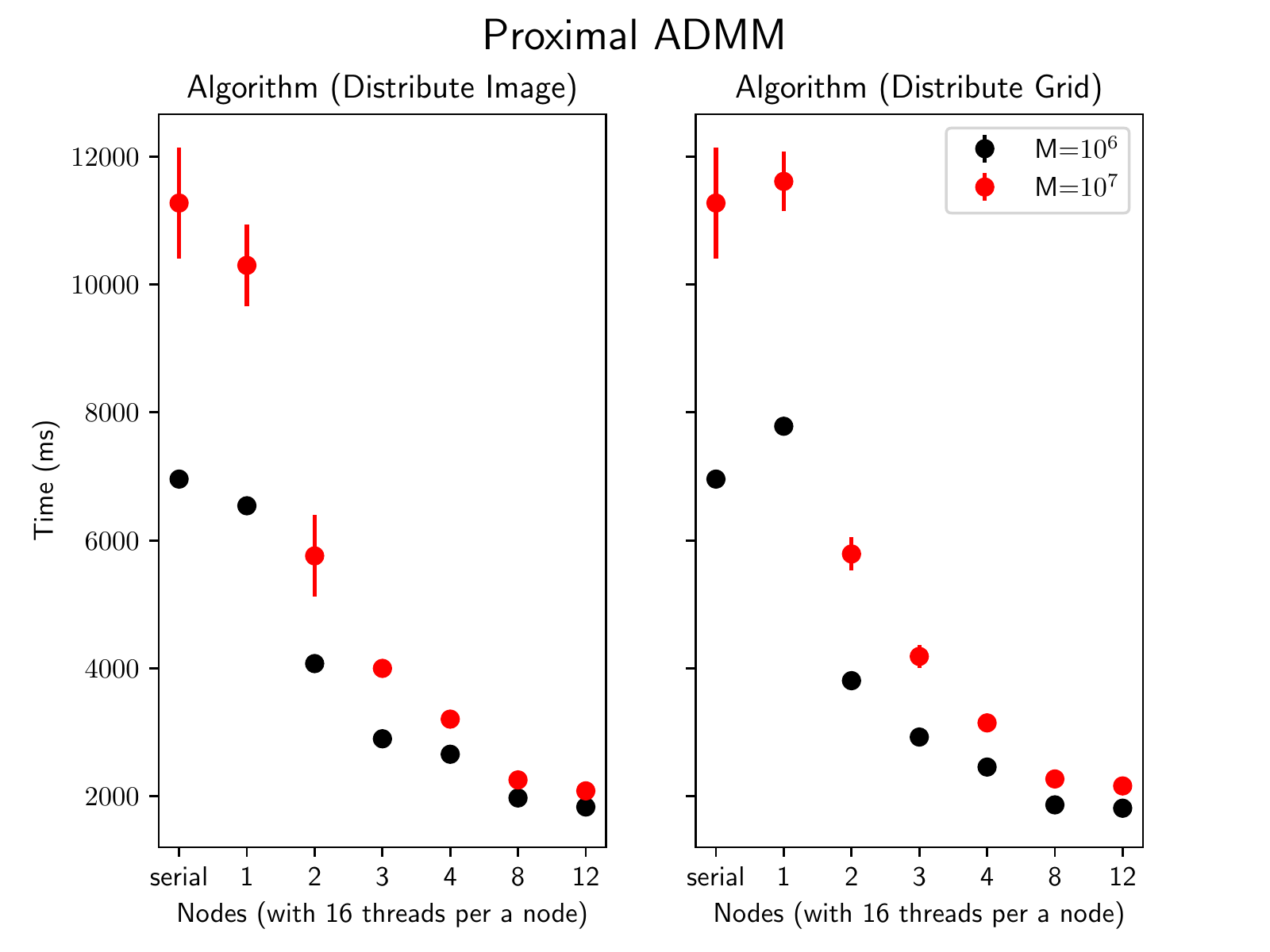}
	\caption{Time to apply a single iteration of the ADMM algorithm as a function of the number of MPI nodes, benchmarked against the non MPI (serial) implementation. The Dual Forward Backward algorithm is limited to one iteration. We fix the number of visibilities and image size at $N = 1024 \times 1024$, $M \in \{10^6,10^7\}$. On the left the MPI implementation corresponds to using Algorithm \ref{alg-admm-image} (which uses the MPI measurement operator from Section \ref{sec:mo_dist_image} where the image is distributed); on the right MPI implementation corresponds to using Algorithm \ref{alg-admm-grid} (which uses the MPI measurement operator from Section \ref{sec:mo_dist_grid} where the Fourier grid is distributed).}
	\label{fig:admm_bench}
\end{figure}

In Figure \ref{fig:admm_bench}, we time the application of one iteration of ADMM using one Dual Forward-Backward iteration. We find a clear increase in performance when increasing the number of nodes used. This is predicted from the performance improvements from the previous sections.

\subsection{GPU Measurement Operator Benchmarks}
\begin{figure}
	\center
	\includegraphics[width=0.5\textwidth]{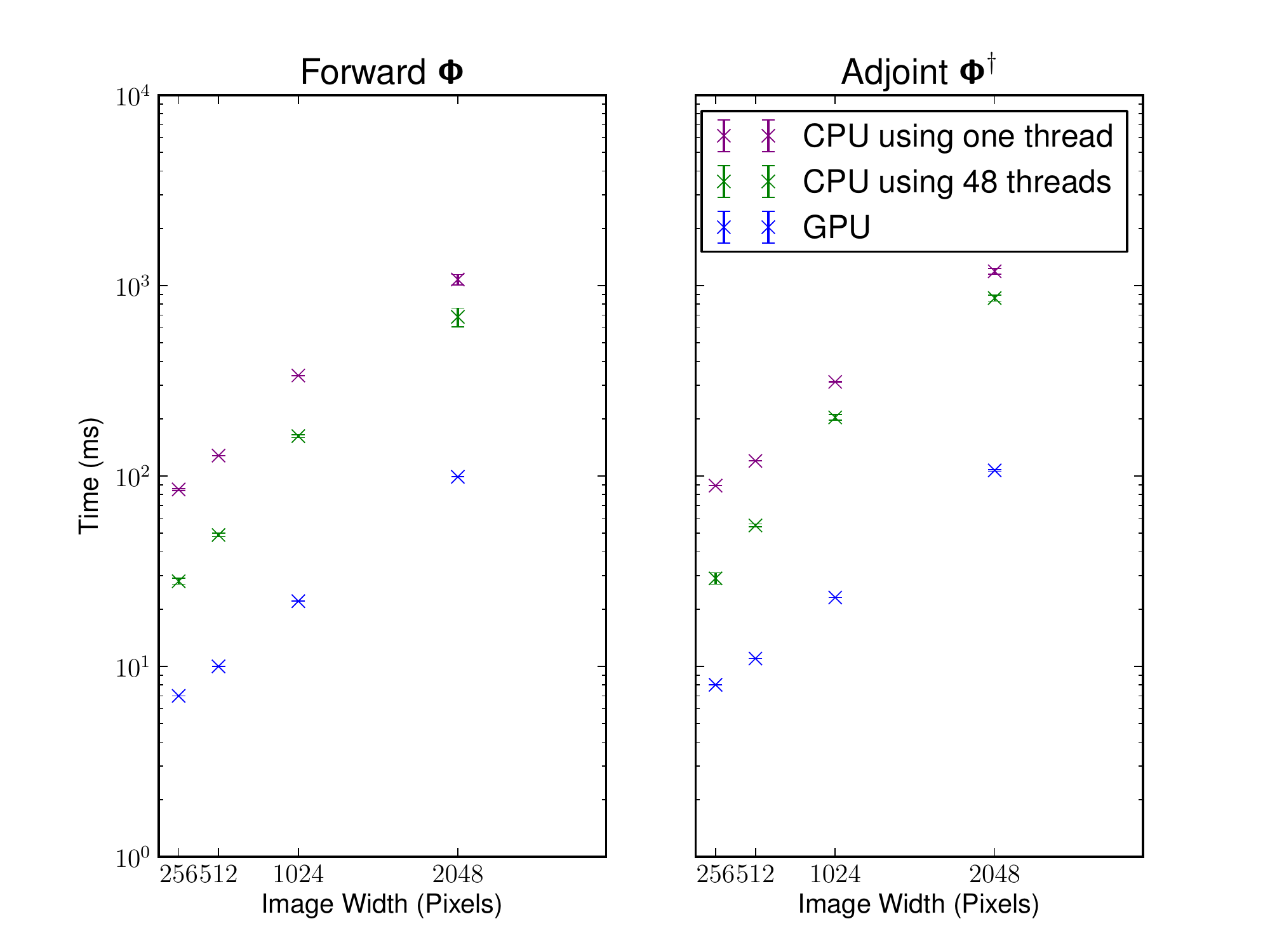}
	\caption{Time to apply forward $\bm{\mathsf{\Phi}}$ and adjoint $\bm{\mathsf{\Phi}}^\dagger$ as a function of image size, using CPU implementation and ArrayFire with GPU CUDA back-end implementation. We fix the number of visibilities at $M = 5 \times 10^6$, and vary the width of a square image. The CPU times for 1 and 48 threads show that there is some improvement by using threading for the CPU. However, it is clear that GPU implementation remains almost an order of magnitude faster for both gridding and degridding, especially at larger image sizes.}
	\label{fig:cpu_gpu_mo_bench}
\end{figure}

The MPI measurement operators in the previous subsection can also make use of graphics processing units (GPUs) to increase performance. We have implemented the MPI measurement operators using the software package ArrayFire \cite{Yalamanchili2015}, which provides the flexibility to chose a CPU, CUDA, or OpenCL back-end to perform computations. The hybrid MPI-GPU measurement operator works the same as the MPI measurement operator, but all operations on a given node are performed on a GPU. In this section, we show that the GPU can increase performance. We benchmark the ArrayFire implementation using a CUDA back-end, against the equivalent measurement operator. No MPI is used in these benchmarks, since it is clear from the previous section that MPI will also increase performance. We perform the benchmarks on a high performance workstation, using an NVIDIA Quadro K4200 GPU (with 4GB RAM). We use $5 \times 10^6$ visibilities, and use the image sizes of $256\times 256$, $512\times 512$, $1024\times 1024$ and $2048\times 2048$. We find that the GPU implementation of degridding and gridding is about 10 times faster than the CPU counter part. Figure \ref{fig:cpu_gpu_mo_bench} shows the application of $\bm{\mathsf{\Phi}}$ and $\bm{\mathsf{\Phi}}^\dagger$, with a large performance improvement when using the GPU. 

{
\section{Application to Big-Data}
\label{sec:big_data}
In the previous section, the benchmarks against the serial version of the algorithm show that the distributed version of the algorithms can in principle increase the run-time speed. In this section, we pre-sample the gridding kernel onto a fine one dimensional grid to compress the storage of the gridding matrix. Since many of the kernel coefficients are reused, this can greatly reduce the memory needed during operation. We then use the pre-sampled kernel to time the distributed measurement operator and ADMM algorithms for larger data sets.

\subsection{Kernel Pre-Sampling}
The amount of memory required to store the sparse matrix $\bm{\mathsf{G}}$ can be quite substantial. When stored in the most naive way, it will take up $J^2M$ times the number of coefficients. This number increases by a factor of 2 if $\bm{\mathsf{G}}^\dagger$ is stored separately to ensure row major order during application. In this section, we discuss our strategy for compressing $\bm{\mathsf{G}}$ by using pre-sampling of the gridding kernel $g(u_{\rm pix})$ onto a fine grid over its support. Zeroth order interpolation can then be applied to calculate $\bm{\mathsf{G}}$  and $\bm{\mathsf{G}}^\dagger$ during application. When $J = 4$ we find that we need to store 16 times as many weights as we do visibilities, which can cause us reach a memory bottleneck due to the limits of memory of the total cluster. It is easy to see that for $J = 4$ and $M = 100 \times 10^6$ will take up at least 25.6 Gb in complex valued weights, and this will double if $\bm{\mathsf{G}}^\dagger$ is stored separately. 

On the fly calculation evaluation of $g(u_{\rm pix})$ frees up memory, but will take up calculation time during the application of $\bm{\mathsf{G}}$. We have found that on the fly calculation time can be reduced by pre-sampling the kernel calculations.

This is not the first time that the idea of kernel pre-sampling has been used. The work of \cite{bea05} discusses how pre-sampling can be used to accurately evaluate gridding kernels. In particular, they investigate the use of zero and first order interpolation. 

In our work, we pre-sample $g(u_{\rm pix})$ for values between $0 \leq |u_{\rm pix}| \leq J/2$. We use the knowledge that $g(u_{\rm pix}) = g(-u_{\rm pix})$, which halves the number of pre-samples needed. We have found that pre-sampling $g(u_{\rm pix})$ to a density of $2 \times 10^{5}$ samples per pixel is sufficient to make zeroth order interpolation accurate to $10^{-5}$. This means that for $J  = 4$, we need $4 \times 10^5$ samples with the advantage that the number of pre-samples is independent of the number of measurements $M$. Furthermore, we can use the same pre-samples for the adjoint operator $\bm{\mathsf{G}}^\dagger$. 

In the next sections, we show the application of the distributed measurement operator and ADMM algorithms to larger data sets. In particular, pre-sampling the kernel calculation unlocks more computational resources during reconstruction of large data sets.

\subsection{Measurement Operator and ADMM Scaling}
Section \ref{sec:benchmarks} shows benchmarks against the serial algorithm, which shows that the distributed algorithms reduce the run-time (by increasing the available computational resources). However, benchmarking against a serial version is not possible or useful for understanding how the performance scales for very large data sets. We pre-sample the gridding kernels in order to time the application of the distributed algorithms to larger data sets.

To measure the timings, we fix the image size to $N = 1024\times 1024$ pixels and vary the number of visibilities per node $M_j$. The timings use the Grace computing cluster at University College London. Each node of Grace contains two 8 core Intel Xeon E5-2630v3 processors (16 cores total) and 64 Gb of RAM.\footnote{More details can be found at \url{https://wiki.rc.ucl.ac.uk/wiki/RC_Systems\#Grace_technical_specs}} It is important to note that the $uv$ sampling pattern has no clear impact on the time per iteration, especially with large data sizes.

Figures \ref{fig:mo_scaling} (measurement operator) and \ref{fig:padmm_scaling} (ADMM) show how the application run-time varies as a function of visibilities per node $M_j$, using $n_{\rm d} = 25, 50, 100$ nodes. We did not find a large impact on performance when increasing the number of nodes for the different types of measurement operators and ADMM algorithms. {This is important, because it suggests that communication overhead and lag in computation across nodes is currently not a major limitation or factor for the implementation of this algorithm up to 100 nodes.}

Figure \ref{fig:padmm_scaling} also shows the total size of the data sets in memory across the cluster as a function of the number of nodes and visibilities per node. We define the total data set as number of complex valued visibilities $\bm{y}_{k}$, the complex valued weights $\bm{\mathsf{W}}_{kk}$, and real valued $\bm{u}_{k}$ and $\bm{v}_{k}$ coordinates. We are using the data types `double' (8 bytes) and `complex double' (16 bytes), this can be multiplied by $M$ or $M_j$ to calculate the total memory over the cluster or per node respectively. The run-times are measured for data sets the ranging from 1 Gb to 2.4 Tb. The largest data sets contain 12.5 billion, 25 billion and 50 billion visibilities, with $M_j = 500 \times 10^6$ visibilities per node. The run-times for ADMM range from 9 seconds to 3 minutes per iteration, with a change in behavior for $M_j > 10^7$. However, the run-times for the measurement operators range from 100 milliseconds to 40 seconds per iteration.

\begin{figure}
	\center
	\includegraphics[width=0.5\textwidth]{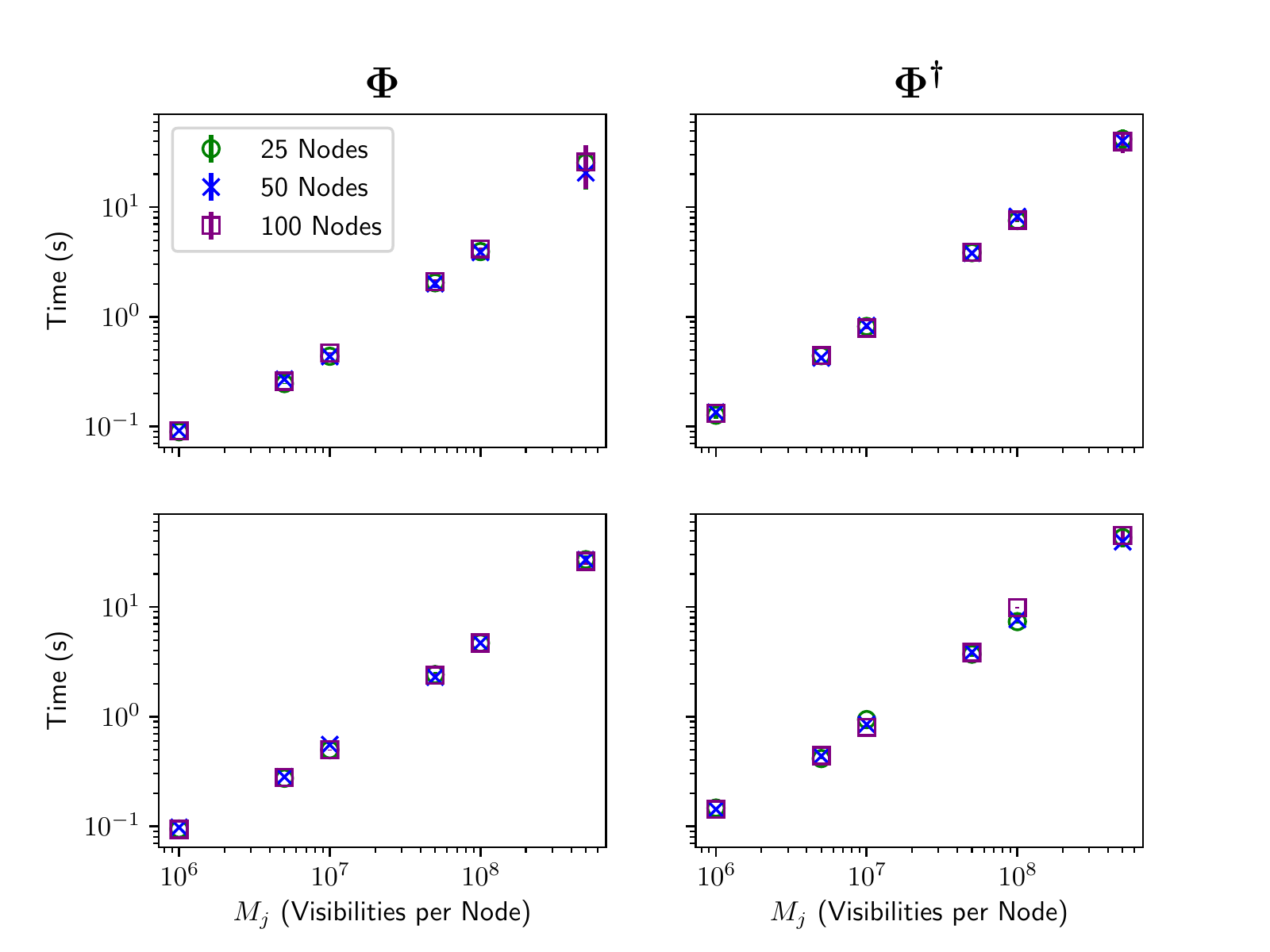}
	\caption{Application times of the distributed image (top row) and distributed grid measurement operators (left column) and their adjoint (right column) as a function of visibilities per node $M_j$. We show the application times for $n_{\rm d} = 25, 50, 100$ nodes. We find that application time increases linearly as a function of $M_j$. Importantly, choosing $n_{\rm d} = 100$ does not have a big impact on application time. The run-times in this figure measured for data sets the ranging from 1 Gb to 2.4 Tb.}
	\label{fig:mo_scaling}
\end{figure}

\begin{figure}
	\center
	\includegraphics[width=0.5\textwidth]{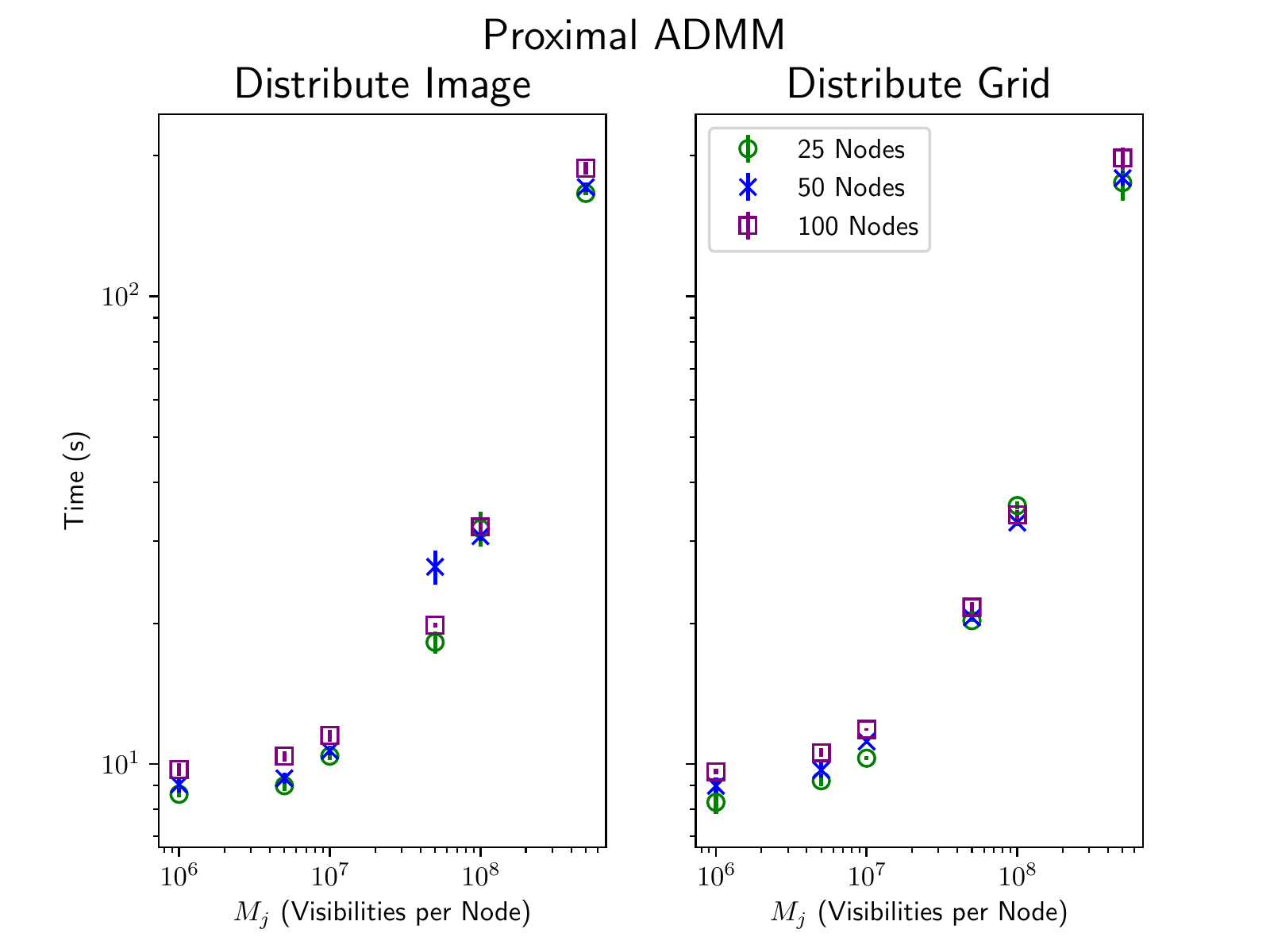}
	\includegraphics[width=0.5\textwidth]{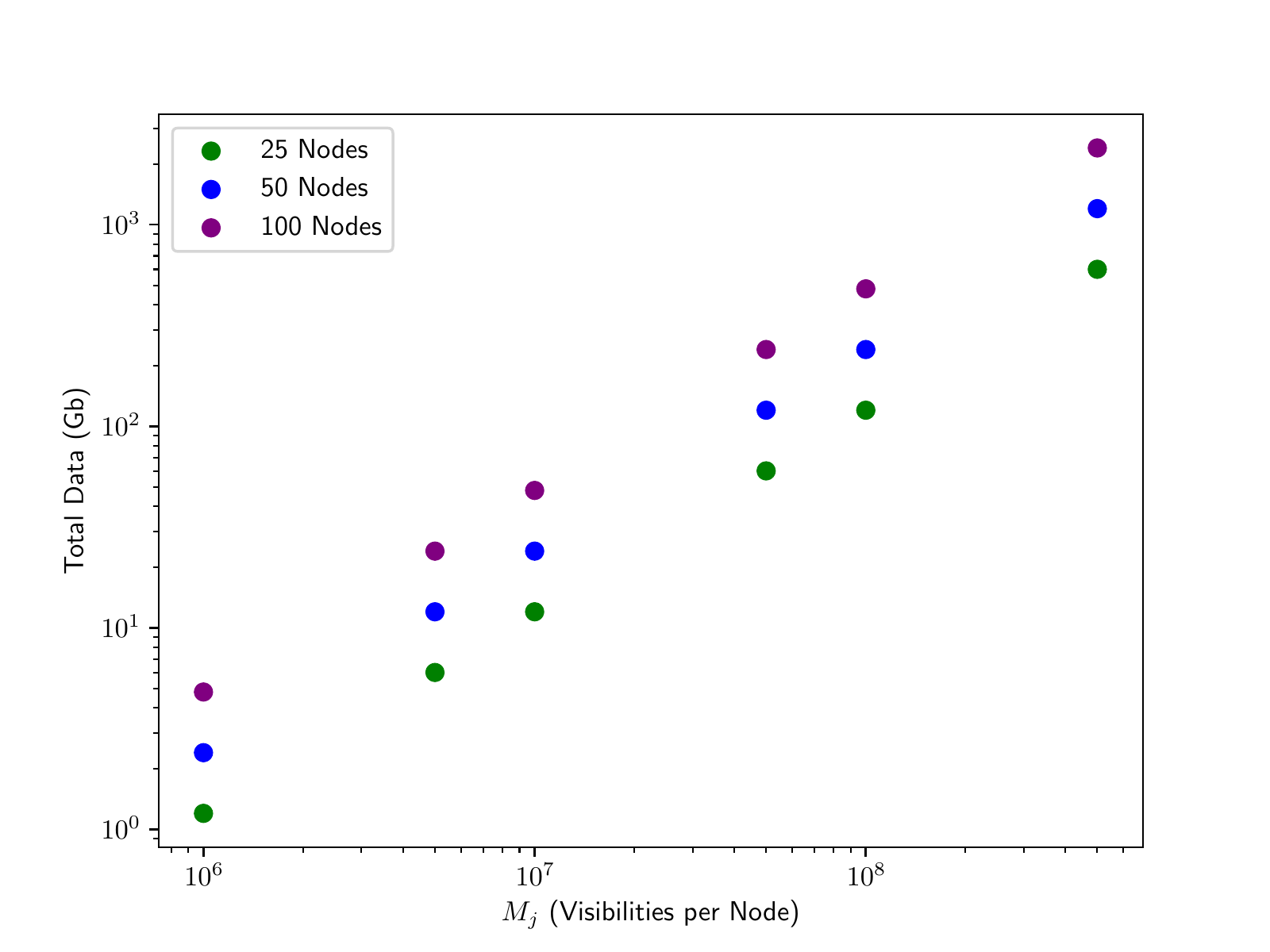}
	\caption{(top) Application times of the distributed image (left) and distributed grid ADMM algorithms (right) as a function of visibilities per node $M_j$. We show the application times for $n_{\rm d} = 25, 50, 100$ nodes. When $M_j > 10^7$, we find that scaling of ADMM changes in how it increases with $M_j$. (Bottom) The total size of data sets used in Figures \ref{fig:mo_scaling} and \ref{fig:padmm_scaling}. The data sets range from 1 Gb to 2.4 Tb.}
	\label{fig:padmm_scaling}
\end{figure}
}
\section{Conclusion}
\label{sec:conclusion}
{
We presented the first implementation of a distributed convex optimization algorithm that uses MPI to reconstruct images from radio interferometric telescopes We have described new algorithms and their application to convex optimization methods for distributed interferometric image reconstruction as implemented by the PURIFY 3.1.0 and SOPT 3.1.0 software packages \cite{pra19,pra19b}. We developed a number of alternative algorithms to better address practical considerations for MPI and include explicit descriptions of the MPI processes used. We then benchmark the distributed implementations to demonstrate considerable computational gains compared to the serial equivalents. We found that the pre-sampled gridding kernel calculation allows the distributed gridding operations and the ADMM algorithm to scale to big data sets that are over 1 Tb while providing a computational speed per iteration that does not have a strong communication overhead for up to 100 nodes. While we plan to make improvements to performance in the future, this work is a significant step towards new interferometric image reconstruction algorithms that can be applied to large data sets.

With next generation radio interferometric telescopes coming online, distributed and parallel image reconstruction and data analysis will be necessary to deal with the large image sizes and large volumes of data of forthcoming telescopes. This work is an important step on the path to developing computational algorithms that will be required for telescopes to reach the high resolution and sensitivity needed for science goals of telescopes such as the SKA.
}
\section*{Acknowledgements}
The authors thank Rafael Carrillo, Alex Onose and Yves Wiaux for useful discussions regarding  the algorithms presented in \cite{ono16}. This work was also supported by the Engineering and Physical Sciences Research Council (EPSRC) through grant EP/M011089/1 and by the Leverhulme Trust.  The authors acknowledge the use of the UCL Legion High Performance Computing Facility (Legion@UCL), and associated support services, in the completion of this work. The authors acknowledge the use of the UCL Grace High Performance Computing Facility (Grace@UCL), and associated support services, in the completion of this work.

 \appendix

\section{Proximal Algorithms}
\label{sec:prox_algo}
Let $X = \mathbb{R}^N, f \in \Gamma_0(X)$, $g \in \Gamma_0(X)$, using the tools from proximal calculus, 
we can solve the convex optimization problem with the general form 
\begin{equation} \label{eqn:model-general}
	\min_{\bm{x} \in \mathbb{R}^N} f(\bm{x}) + g(\bm{x})\, .
\end{equation}
Here, for simplicity, we assume each of the minimization problems, like \eqref{eqn:model-general} considered in this paper has a global minimizer.
If the proximal operator of $f + g$ was known or could be computed easily, we could recursively iterate the proximal operator to find a solution to \eqref{eqn:model-general}. 
However, we often only know the proximal operator for $f$ and $g$ separately.  
In the following, we briefly introduce a few algorithms among the proximal algorithm category which can address this kind of minimization problem. Moreover, these algorithms can be adapted to be distributed across computing clusters \cite{com09,boy11,kom15,ono16}; one example is used later in this work.

\subsection{Forward-Backward Splitting} \label{sec-fb}
In the case that $f$ is differentiable, problem \eqref{eqn:model-general} can be solved using
the Forward-Backward splitting algorithm. Starting with a proper initialization, $\forall \lambda \in (0, +\infty)$, the iterative scheme can be represented as
\begin{equation} \label{eqn:alg-fb}
	\bm{x}^{(k+1)} = {\rm prox}_{\lambda g}(\bm{x}^{(k)} - \lambda \nabla f(\bm{x}^{(k)})),
\end{equation}
which includes a forward gradient step (explicit) regarding function $f$ and a backward step (implicit and involves solving a 
proximal operator) with respect to $g$. Refer to \cite{dau04,bec09,gir15,gar15,kom15,ono16} and references therein for more details and the variants of the Forward-Backward splitting algorithm.

As an example, we see that formula \eqref{eqn:alg-fb} can be directly used to solve the unconstrained problem \eqref{eq:MAP-estimate}
so as to obtain an MAP estimator of the sky in radio astronomy. When $g$ is the $\ell_1$-norm, this algorithm becomes the Iterative Shrinkage-Thresholding Algorithm (ISTA), where it is possible to obtain accelerated convergence by using Fast ISTA (FISTA) \cite{bec09}, which is detailed in Algorithm \ref{alg:fista}.
\begin{algorithm}[ht]
	\caption{FISTA}
	\label{alg:fista}			  
	\begin{algorithmic}[1]
		\small
		\Given{$\bm{x}^{(0)} \in \mathbb{R}^N, \lambda > 0, \theta_0 = 1, \hat{\bm{x}}^{(0)} = \bm{x}^{(0)}$}
		\RepeatFor{$k=0,\ldots$}
		\State $\bm{x}^{(k+1)} = {\rm prox}_{\lambda g}(\hat{\bm{x}}^{(k)} - \lambda \nabla f(\hat{\bm{x}}^{(k)}))$
		\State $\theta_{k + 1} = \frac{1 + \sqrt{1 + 4\theta_k^2}}{2}$
		\State  $\hat{\bm{x}}^{(k+1)} = \bm{x}^{(k+1)} + \frac{\theta_k - 1}{\theta_{k+1}} (\bm{x}^{(k+1)} - \bm{x}^{(k)})$ 
		\Until {\bf convergence \normalfont}
	\end{algorithmic}
\end{algorithm}

The Forward-Backward algorithm is often simpler to compute than the algorithms that follow, which is an advantage of solving the unconstrained problem over the constrained problem. However, there are many cases where $f$ is not differentiable; for example when it represents an indicator function. Note that the Forward-Backward algorithm cannot be used to solve the constrained problem \eqref{l1-constrained} directly, due to the non-differentiable indicator function.
%

\subsection{Douglas-Rachford Splitting}
When both $f$ and $g$ in \eqref{eqn:model-general} are non-differentiable, the Douglas-Rachford splitting algorithm can be applied; see \cite{com07,bot13} for more details on the Douglas-Rachford splitting algorithm. 
Its iterative formula, $\forall \lambda \in (0, +\infty)$, reads 
\begin{align}
\begin{split}
\begin{cases}
\bm{x}^{(k)} = {\rm prox}_{\lambda g}(\bm{v}^{(k)}), &  \\
\bm{v}^{(k+1)} = \bm{v}^{(k)} + \gamma^{(k)} ({\rm prox}_{\lambda f}(2\bm{x}^{(k)} - \bm{v}^{(k)}) - \bm{x}^{(k)}),
\end{cases}
\end{split}
\end{align}
where $\gamma^{(k)} \in (\alpha, 2-\alpha)$, $\alpha \in (0, 1)$. This iterative scheme needs the proximal operator for $f$ and $g$ individually.
Therefore, the Douglas-Rachford splitting algorithm is restricted by the degree of difficulty of computing the proximal operators of $f$ and $g$. The algorithm is summarized in Algorithm \ref{alg:dr}.

As an example, the Douglas-Rachford splitting algorithm can theoretically be used to solve the constrained problem \eqref{eq:model-con} after moving its 
constraint into the objective functional by using the indicator function on an $\ell_2$-ball. However, if $\bm{\mathsf{\Phi}}$ is not an identity operator, as in radio interferometry, solving the proximal operator of this kind of indicator function is not easy computationally.

\begin{algorithm}[ht]
	\caption{Douglas-Rachford Splitting Algorithm}
	\label{alg:dr}			  
	\begin{algorithmic}[1]
		\small
		\Given{$\bm{v}^{(0)} \in \mathbb{R}^N, \alpha \in (0, 1), \lambda > 0$}
		\RepeatFor{$k=0,\ldots$}
		\State $\bm{x}^{(k)} = {\rm prox}_{\lambda g}(\bm{v}^{(k)})$
		\State  $\gamma^{(k)} \in (\alpha, 2-\alpha)$
		\State $\bm{v}^{(k+1)} = \bm{v}^{(k)} + \gamma^{(k)} ({\rm prox}_{\lambda f}(2\bm{x}^{(k)} - \bm{v}^{(k)}) - \bm{x}^{(k)})$
		\Until {\bf convergence \normalfont}
	\end{algorithmic}
\end{algorithm}

\subsection{Alternating Direction Method of Multipliers}
\label{sec-admm}
The Forward-Backward and Douglas-Rachford splitting algorithms presented above require the proximal operators $f$ and $g$ 
to be easy to compute. In practice, this is sometimes not the case. For example, when function $f$ involves explicitly 
a linear transformation $\bm{\mathsf{L}} \in \mathbb{R}^{K\times N}$ (e.g. a measurement operator), we must consider the problem
\begin{equation} \label{eqn:model-general-L}
	\min_{\bm{x} \in \mathbb{R}^N} f(\bm{\mathsf{L}} \bm{x}) + g(\bm{x}),
\end{equation}
where the proximal operator of $f(\bm{\mathsf{L}} \bm{x})$ has no explicit expression.

Problem \eqref{eqn:model-general-L} can be addressed by the alternating direction method of multipliers (ADMM) \cite{boy11,yan11,ono16}. 
After setting $\bm{v} = \bm{\mathsf{L}}\bm{x}$, problem \eqref{eqn:model-general-L} becomes
\begin{equation} \label{eqn:model-general-L-con}
	\min_{\bm{x} \in \mathbb{R}^N} f(\bm{v}) + g(\bm{x}), \quad {\rm s.t.} \quad \bm{v} = \bm{\mathsf{L}}\bm{x}.
\end{equation}
This problem has the following augmented Lagrangian with index $\lambda \in (0, +\infty)$
\begin{equation}
	\mathcal{L}(\bm{x}, \bm{v}, \bm{z})
		:= f(\bm{v}) + g(\bm{x}) + \frac{1}{\lambda}\bm{z}^\dagger (\bm{\mathsf{L}} \bm{x} - \bm{v}) 
		+ \frac{1}{2\lambda}\| \bm{\mathsf{L}} \bm{x} - \bm{v} \|_{\ell_2}^2,
\end{equation}
which can be solved alternatively corresponding to $\bm{x}, \bm{v}, \bm{z}$. More precisely, $\mathcal{L}$ is minimized with respect to variables $\bm{x}$ and $\bm{v}$ alternatively while updating the dual variable 
$\bm{z}$ (using the dual ascent method \cite{boy11}) to ensure that the constraint $\bm{v} = \bm{\mathsf{L}}\bm{x}$ is met in the final solution, i.e.,
\begin{align}
		\bm{x}^{(k)} & = \argmin_{\bm{x}\in \mathbb{R}^N} \mathcal{L}(\bm{x}, \bm{v}^{(k)}, \bm{z}^{(k)}), \\
		\bm{v}^{(k+1)} & = \argmin_{\bm{v}\in \mathbb{R}^K} \mathcal{L}(\bm{x}^{(k)}, \bm{v}, \bm{z}^{(k)}), \\
		{\bm{z}}^{(k+1)} & = \bm{z}^{(k)} +  (\bm{\mathsf{L}}\bm{x}^{(k)} - \bm{v}^{(k+1)}),
\end{align}
which can be rewritten as
\begin{align}
		\bm{x}^{(k)} & = \argmin_{\bm{x}\in \mathbb{R}^N} \left(g(\bm{x}) + \frac{1}{2\lambda}\|\bm{\mathsf{L}}\bm{x} - (\bm{v}^{(k)} - \bm{z}^{(k)})\|^2_{\ell_2}\right), \label{alg:admm-x}\\
		\bm{v}^{(k+1)} & = \argmin_{\bm{v}\in \mathbb{R}^K} \left(f(\bm{v}) + \frac{1}{2\lambda}\|\bm{v} -(\bm{\mathsf{L}}\bm{x}^{(k)} + \bm{z}^{(k)})\|^2_{\ell_2}\right), \label{alg:admm-v} \\
		{\bm{z}}^{(k+1)} & = \bm{z}^{(k)} +  (\bm{\mathsf{L}}\bm{x}^{(k)} - \bm{v}^{(k+1)}).
\end{align}
Note, importantly, that the above problem \eqref{alg:admm-v} is actually computing the proximal operator of function $f$ without involving the operator $\bm{\mathsf{L}}$, which circumvents computing the proximal operator of $f(\bm{\mathsf{L}} \bm{x})$ directly and generally has an explicit expression.
We comment that ADMM has a close relationship to the Douglas-Rachford algorithm, see \cite{com09,kom15} for more details.
The procedures of ADMM are briefly summarized in Algorithm \ref{alg-padmm}, where we define
\begin{equation}
{\rm prox}_{\lambda g}^{\bm{\mathsf{L}}} (\bm{u}) = \argmin_{\bm{x}\in \mathbb{R}^N} \left(g(\bm{x}) + \frac{1}{2\lambda}\|\bm{\mathsf{L}}\bm{x} - \bm{u}\|_{\ell_2}^2\right) \, ,
\end{equation}
which may have a simple closed-form solution, or can be solved iteratively using a Forward-Backward method since its second term is differentiable.

\begin{algorithm}[ht]
	\caption{Alternating Direction Method of Multipliers (ADMM)}
	\label{alg-padmm}			  
	\begin{algorithmic}[1]
		\small
		\Given{$\bm{z}^{(0)},\bm{v}^{(0)} \in \mathbb{R}^K, \lambda > 0$}
		\RepeatFor{$k=0,\ldots$}
		\State $\bm{x}^{(k)} = {\rm prox}_{\lambda g}^{\bm{\mathsf{L}}}(\bm{v}^{(k)} - \bm{z}^{(k)})$
		\State $\bm{v}^{(k+1)} = {\rm prox}_{\lambda f}(\bm{\mathsf{L}}\bm{x}^{(k)} + \bm{z}^{(k)})$
		\State $\bm{z}^{(k+1)} = \bm{z}^{(k)} + \bm{\mathsf{L}}\bm{x}^{(k)} - \bm{v}^{(k+1)}$
		\Until {\bf convergence \normalfont}
	\end{algorithmic}
\end{algorithm}

A generalization of ADMM is simultaneous direction method of multipliers (SDMM), which can be applied to an objective function of more than two functions \cite{set10,car14}. However, this method often requires operator inversion which can be expensive \cite{ono16}.

In this work, after setting $f(\bm{\mathsf{L}}\bm{x})$ to be the indicator function on an $\ell_2$-ball, we can use ADMM to solve the constrained problem \eqref{eq:model-con} with the positivity constraint, i.e. the problem \eqref{l1-constrained}. 
We approximately solve ${\rm prox}_{\lambda g}^{\bm{\mathsf{L}}} (\bm{u})$ using an iteration of the Forward-Backward splitting method, and then use the Dual Forward-Backward splitting algorithm (which will be presented in Section \ref{sec-dual-fb}) to solve ${\rm prox}_{\lambda g}(\bm{u})$ iteratively, where $g$ contains the $\ell_1$-norm and the positivity constraint (see \cite{ono16} and Section \ref{sec:ri_padmm} for more detail).
\subsection{Primal-Dual Splitting}
\label{sec-primaldual}
In addition to ADMM, problem \eqref{eqn:model-general-L} can also be solved by the Primal-Dual splitting algorithm; an algorithm that like ADMM can be adapted to be distributed and performed in parallel \cite{kom15,ono16}. 
Firstly, the primal problem \eqref{eqn:model-general-L} can be rewritten as the following 
Primal-Dual formulation, i.e.,
\begin{equation} \label{eqn:model-pd-L}
	\min_{\bm{x} } \max_{\bm{z}} g(\bm{x}) + \langle \bm{\mathsf{L}}\bm{x}, \bm{z} \rangle  -  f^*(\bm{z}),
\end{equation}
which is a saddle point problem, where $\langle \bm{\mathsf{L}}\bm{x}, \bm{z} \rangle  = \bm{z}^\dagger\bm{\mathsf{L}}\bm{x}$. 
It can be solved from minimizing and maximizing with respect to $\bm{x}$ and ${\bm{z}}$ alternatively,
where for each subproblem the Forward-Backward ideas presented in Section \ref{sec-fb} can be applied if needed. The Primal-Dual algorithm 
is summarized in Algorithm \ref{alg-pd}.  Furthermore, Moreau decomposition in equation \eqref{eq:Moreau_decomposition} can be used to calculate the proximal operator of $f^*$ given the proximal operator of $f$, i.e. \[ {\rm prox}_{\sigma f^*}(\bm{z}) = \bm{z} - \sigma {\rm prox}_{\sigma^{-1}f}(\bm{z}/\sigma )\, .\]

The Primal-Dual and ADMM algorithms are both very efficient algorithms to solve problems like \eqref{eqn:model-general-L}.
The Primal-Dual algorithm generally can achieve better convergence rates than ADMM. 
However, since ADMM needs to compute the proximal operators 
${\rm prox}_{\lambda f}$ and ${\rm prox}_{\lambda g}^{\bm{\mathsf{L}}}$ and the Primal-Dual algorithm needs to compute ${\rm prox}_{\sigma f^*}$ and ${\rm prox}_{\tau g}$, 
which method is more appropriate often depends on the overall problem itself. 
In addition, there has been plenty of work to optimize them further, which makes their performance more comparable
and in some cases equivalent to each other (see \cite{kom15} for an overview of Primal-Dual methods).

\begin{algorithm}[ht]
	\caption{Primal-Dual Algorithm}
	\label{alg-pd}			  
	\begin{algorithmic}[1]
		\small
		\Given{$\bm{x}^{(0)} \in \mathbb{R}^N, \bm{z}^{(0)} \in \mathbb{R}^N, \tau, \sigma > 0, \theta \in [0, 1]$}
		\RepeatFor{$k=0,\ldots$}
		\State $\bm{z}^{(k+1)} = {\rm prox}_{\sigma f^*}(\bm{z}^{(k)} + \sigma \bm{\mathsf{L}} \hat{\bm{x}}^{(k)})$
		\State $\bm{x}^{(k+1)} = {\rm prox}_{\tau g}(\bm{x}^{(k)} - \tau \bm{\mathsf{L}}^{\dagger} \bm{z}^{(k+1)})$
		\State $\hat{\bm{x}}^{(k+1)} = \bm{x}^{(k+1)} + \theta(\bm{x}^{(k+1)} - \bm{x}^{(k)})$
		\Until {\bf convergence \normalfont}
	\end{algorithmic}
\end{algorithm}

\subsection{Dual Forward-Backward Splitting}
\label{sec-dual-fb}
An algorithm closely related to the Primal-Dual algorithm is known as the Dual Forward-Backward splitting algorithm \cite{com10,kom15}. 
To obtain the dual problem of \eqref{eqn:model-general-L}, using Lagrangian multiplier $\bm{z}$, we get the Lagrangian
\begin{equation}
	\mathcal{L}(\bm{x}, \bm{v}, \bm{z})
		:= f(\bm{v}) + g(\bm{x}) + \langle \bm{\mathsf{L}}\bm{x} -\bm{v}, \bm{z} \rangle\, .
\end{equation}
By minimizing the Lagrangian over $\bm{x}$ and ${\bm{v}}$, we have
\begin{align}
\begin{split}
	\inf_{\bm{x}, \bm{v}} \mathcal{L}(\bm{x}, \bm{v}, \bm{z}) &= 
		- \sup_{\bm{v}}\left (\langle \bm{z}, \bm{v}\rangle - f(\bm{v}) \right)
		-\sup_{\bm{x}}\left (\langle -\bm{\mathsf{L}}^\dagger\bm{z}, \bm{x}\rangle - g(\bm{x}) \right)  \\
		&= - f^*(\bm{z}) - g^*( -\bm{\mathsf{L}}^\dagger\bm{z}).
\end{split}		
\end{align}
Then we have the dual problem of problem \eqref{eqn:model-general-L}, i.e.
\begin{equation} \label{eqn:fb-dual}
	\min_{\bm{z}} f^*(\bm{z}) + g^*( -\bm{\mathsf{L}}^\dagger\bm{z}).
\end{equation}

Note that term $g^*( -\bm{\mathsf{L}}^\dagger\bm{z})$ is differentiable and it is shown in \cite{boy15} that
\begin{equation}  \label{eqn:fb-dual-g}
	\partial_{\bm{z}} g^* = -\bm{\mathsf{L}} \left (\argmin_{\bm{v}\in X} \left \{\langle\bm{\mathsf{L}}^\dagger\bm{z}, \bm{v}\rangle  + g(\bm{v})\right\} \right)\, .
\end{equation}
Let $\bar{g}(\bm{z}) = g^*( -\bm{\mathsf{L}}^\dagger\bm{z})$,
applying the Forward-Backward splitting iterative scheme \eqref{eqn:alg-fb} with the relaxation
on \eqref{eqn:fb-dual} $f^* + g^* \to \sigma f^* + \sigma g^*$ with $\sigma \in (0, \infty)$ and combing with the dual ascent method \cite{boy11}, we have 
\begin{align}
	\bm{z}^{(k+1)} & = {\rm prox}_{\sigma f^*}({\bm{z}}^{(k)} - \sigma \nabla \bar{g}(\hat{\bm{z}}^{(k)})),  \label{eqn:alg-fb-dual-z} \\
	\hat{\bm{z}}^{(k+1)} & = \bm{z}^{(k+1)} + \theta (\bm{z}^{(k+1)} - \bm{z}^{(k)}).  \label{eqn:alg-fb-dual-z1}
\end{align}
which is the so-call Dual Forward-Backward splitting algorithm. 

In particular, applying the Forward-Backward splitting iterative scheme \eqref{eqn:alg-fb} on the minimization problem in \eqref{eqn:fb-dual-g},
we have 
\begin{equation}  \label{eqn:fb-dual-g-fb}
	\nabla \bar{g} (\bm{z}) = -\bm{\mathsf{L}} {\rm prox}_{\tau g} (\bm{x} - \tau \bm{\mathsf{L}}^\dagger\bm{z}).
\end{equation}
Let $\bm{x}^{(k+1)} = {\rm prox}_{\tau g} (\bm{x}^{(k)} - \tau \bm{\mathsf{L}}^\dagger \hat{\bm{z}}^{(k)})$ and substituting \eqref{eqn:fb-dual-g-fb} into \eqref{eqn:alg-fb-dual-z},
we have the following iteration scheme

\begin{align}
		\bm{x}^{(k+1)} & = {\rm prox}_{\tau g} (\bm{x}^{(k)} - \tau \bm{\mathsf{L}}^\dagger \hat{\bm{z}}^{(k)})\\
		\bm{z}^{(k+1)} & = {\rm prox}_{\sigma f^*}({\bm{z}}^{(k)} + \sigma \bm{\mathsf{L}} \bm{x}^{(k+1)})\\
		\hat{\bm{z}}^{(k+1)} & = \bm{z}^{(k+1)} + \theta (\bm{z}^{(k+1)} - \bm{z}^{(k)})
\end{align}
After rearranging the order of the variables and replacing the relaxation strategy for $\bm{z}$ by ${\bm{x}}$,
the above Dual Forward-Backward splitting algorithm turns into the Primal-Dual algorithm (see Algorithm \ref{alg-pd}). 
See \cite{kom15} for more discussions about the relation between the Dual Forward-Backward splitting algorithm and the Primal-Dual algorithm.
\vspace{1cm}



\bibliographystyle{elsarticle-num}
\bibliography{refs}







\end{document}